\documentclass[a4paper,fleqn,usenatbib]{mnras}

\usepackage[T1]{fontenc}
\usepackage{ae,aecompl}

\usepackage{graphicx}	% Including figure files
\usepackage{amsmath}	% Advanced maths commands
\usepackage{amssymb}	% Extra maths symbols
\usepackage{subfigure}
\usepackage{multirow}
\usepackage{xcolor}
\newcommand{\pcc}{\,{\rm cm}^{-3}}

\newcommand{\pcs}{\,{\rm cm}^{-2}}

\newcommand{\kel}{\, {\rm K}}

\newcommand{\msun}{\, {\rm M}_\odot}
\newcommand{\nh}{n_{_{\rm H}}}

\newcommand{\pc}{\, {\rm pc}}

\newcommand{\kpc}{\, {\rm kpc}}

\newcommand{\myr}{\, {\rm Myr}}
\newcommand{\ug}{\, {\rm \mu G}}

\newcommand{\kms}{\, {\rm km \, s^{-1}}}

\title[Molecular cloud line emission]{Do simulated molecular clouds look like real ones?}

\author[Priestley et al.]{
F. D. Priestley\thanks{Email: priestleyf@cardiff.ac.uk}, P. C. Clark \& A. P. Whitworth
\\
School of Physics and Astronomy, Cardiff University, Queen's Buildings, The Parade, Cardiff CF24 3AA, UK \\
}

\date{Accepted XXX. Received YYY; in original form ZZZ}

\pubyear{2022}

\begin{document}
\label{firstpage}
\pagerange{\pageref{firstpage}--\pageref{lastpage}}
\maketitle

\begin{abstract}

  Simulations of molecular clouds often begin from highly idealised initial conditions, such as a uniform-density sphere with an artificially imposed turbulent velocity field. While the resulting structures may appear qualitatively similar to those detected in continuum and line observations, it is unclear whether they are genuinely representative of real molecular clouds. Recent observational work has discovered a tight, often close-to-linear relationship between the integrated intensity of molecular lines and the total column density of the cloud material. We combine magnetohydrodynamical simulations, time-dependent chemistry, and radiative transfer to produce synthetic molecular line observations of model clouds. We find similarly tight correlations between line intensity and column density to those observed, although the linear behaviour is only seen in isolated (as opposed to colliding) model clouds. This linear relationship is not due to optically thin emission; all lines investigated have high optical depths, and the increase in integrated intensity with column density is due to higher velocity dispersion along the line of sight. Overall, the idealised models commonly used in the literature appear to be reasonably accurate representations of real molecular clouds.

\end{abstract}
\begin{keywords}
astrochemistry -- stars: formation -- ISM: molecules -- ISM: clouds -- ISM: structure
\end{keywords}
%%%%%

%%%%%
\section{Introduction}
%%%%%

Stars form in molecular clouds, regions of the interstellar medium (ISM) where densities are high enough ($\nh > 100 \pcc$) and temperatures low enough ($T \sim 10-20 \kel$) for most atoms to be bound in molecules \citep{bergin2007}. How these clouds themselves are formed is not completely understood. Simulations on the scales of entire galaxies can track the conversion of ISM material into the dense, star-forming phase \citep[e.g.][]{dobbs2015,walch2015,duartecabral2016}, but typically lack the resolution to follow the formation of individual stars or stellar systems (although with some exceptions; \citealt{seifried2017b,smith2020,whitworth2022}). Cloud-scale simulations therefore generally start from highly idealised initial conditions, such as spheres or periodic boxes, with artificially-induced turbulence and (if present) uniform magnetic fields \citep[e.g.][]{federrath2016,wurster2019}. We explore here how well these simulated clouds correspond to those actually observed.

A key issue is that simulation outputs do not directly correspond to observable quantities. Most column densities for molecular clouds are derived from thermal dust emission, but the optical properties of interstellar dust are notoriously uncertain \citep{clark2019b,whitworth2019}, and this only provides a two-dimensional view of the structure. Extracting physical properties from molecular line observations has its own complications, including the effects of optical depth, excitation and abundance, and structures seen in PPV space do not always correspond to physical ones in PPP space \citep{clarke2018}. {The relationship between the polarization angle of thermal dust emission and the direction of the magnetic field responsible for it is highly complex, involving the effects of grain alignment, viewing angle, and line-of-sight field variations \citep{chen2016,king2018,seifried2019}. Measurements of the magnetic field strength itself are subject to large uncertainties, both experimental and systematic \citep{lyo2021,skalidis2021,pattle2022}.} Great care should be exercised when comparing the {predicted physical properties of} simulated clouds with {those derived from} observations, as the quantities involved are not measuring the same things.

An alternative approach is to produce synthetic observations of simulated clouds, which can then be compared directly to the data. In addition to the underlying physical model, this requires assumptions about the dust properties and/or the molecular composition of the gas, but the latter at least can be calculated with a model of the time-dependent chemical evolution. Modern simulations often include CO chemistry due to its importance as a coolant, allowing a self-consistent determination of its predicted line intensities \citep[e.g.][]{penaloza2017,seifried2017a,clark2019}, and there are plentiful observational data for various CO transitions and isotopologues \citep[e.g.][]{rigby2019,eden2020,duartecabral2021}. However, high molecular abundances and low critical densities make these lines poor tracers of the high-density regions where stars actually form \citep{shirley2015}. So far, most work on molecules tracing denser material has relied on somewhat ad-hoc assumptions about their abundances \citep{smith2012,smith2013,jones2021}, which may not accurately represent the actual chemical behaviour \citep{priestley2020}.

\citet{tafalla2021} have demonstrated a technique to efficiently sample how individual line intensities vary with total gas column density in molecular clouds, via randomly sampling regions of different column density. Achieving this for species other than CO has previously involved a significant investment of telescope time \citep{pety2017,kauffmann2017,barnes2020}, limiting the number of clouds which could be covered. \citet{tafalla2021} find extremely regular behaviour in the Perseus molecular cloud for most of the lines observed, with intensity increasing almost exactly linearly with column density. This represents an ideal test for models of star formation - if simulated clouds do not show the same behvaiour, then some aspect of the underlying physical/chemical model is incorrect, or at least not applicable to this particular object. In this paper, we use a combination of magnetohydrodynamic (MHD) simulations, time-dependent chemistry, and radiative transfer to predict the intensity-column density relationships of simulated molecular clouds, for a number of commonly-observed lines. By comparing these with the \citet{tafalla2021} data, we investigate whether common model assumptions actually produce clouds which resemble real ones.

\section{Method}

\begin{figure}
  \centering
  \includegraphics[width=\columnwidth]{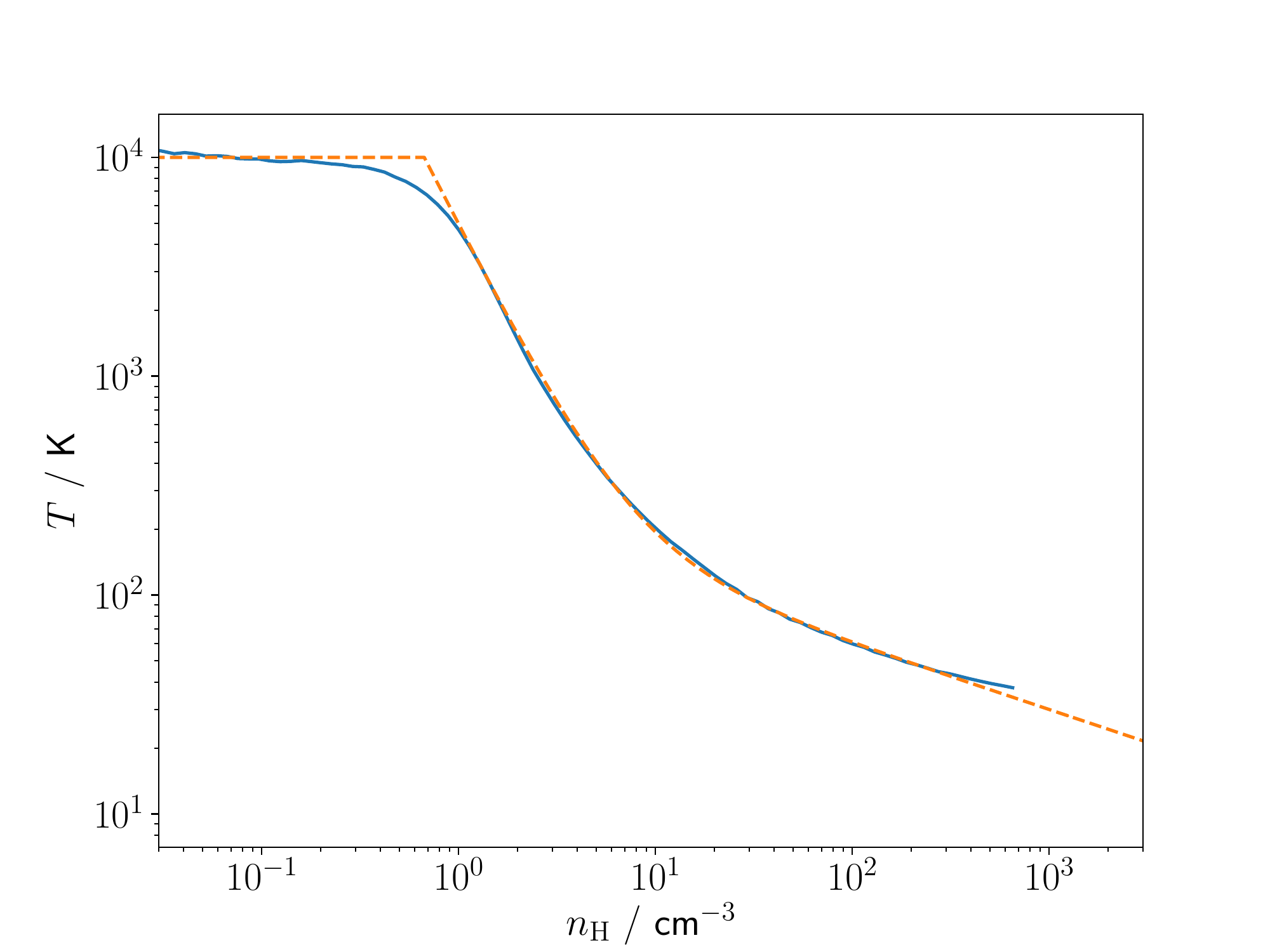}
  \caption{Equilibrium gas temperature as a function of density for a Galactocentric radius of $8.5 \kpc$ and a column density of $10^{19} \pcs$ from \citet{wolfire2003} (blue solid line), and the relation given by Equation \ref{eq:eos} (orange dashed line) with a floor (ceiling) of $10 \kel$ ($10^4 \kel$) imposed.}
  \label{fig:eos}
\end{figure}

\begin{table*}
  \centering
  \caption{Properties of models in this paper: type (isolated or colliding), cloud mass, radius, initial density, turbulent velocity, initial magnetic field strength and direction, collision velocity and cloud offset (for colliding models), and end-point.}
  \begin{tabular}{ccccccccccc}
    \hline
    Model & Type & $M$/$\msun$ & $R$/$\pc$ & $\nh$/$\pcc$ & $\left<v\right>$/$\kms$ & $B_0$/$\ug$ & Dir. & $v_{\rm coll}$/$\kms$ & $\Delta y$/$\pc$ & $t_{\rm end}$/$\myr$ \\
    \hline
    TURB & Isolated & $10^3$ & $3.0$ & $256$ & $1.0$ & $0.0$ & - & - & - & $2.5$ \\
    MAG & Isolated & $10^3$ & $3.0$ & $256$ & $1.0$ & $7.6$ & $z$ & - & - & $2.8$ \\
    COLL & Colliding & $10^3$ & $3.0$ & $256$ & $1.0$ & $0.0$ & - & $2.0$ & $0.0$ & $1.2$ \\
    SHEAR & Colliding & $10^3$ & $3.0$ & $256$ & $1.0$ & $0.0$ & - & $2.0$ & $1.0$ & $1.1$ \\
    COLLBPAR & Colliding & $10^3$ & $3.0$ & $256$ & $1.0$ & $7.6$ & $z$ & $2.0$ & $0.0$ & $1.3$ \\
    COLLBPERP & Colliding & $10^3$ & $3.0$ & $256$ & $1.0$ & $7.6$ & $y$ & $2.0$ & $0.0$ & $2.0$ \\
    \hline
  \end{tabular}
  \label{tab:models}
\end{table*}

\begin{table}
  \centering
  \caption{{Initial gas-phase elemental abundances from \citet{sembach2000}, relative to hydrogen nuclei, used in the chemical modelling.}}
  \begin{tabular}{ccccc}
    \hline
    Element & Abundance & & Element & Abundance \\
    \hline
    C & $1.4 \times 10^{-4}$ & & S & $1.2 \times 10^{-5}$ \\
    N & $7.6 \times 10^{-5}$ & & Si & $1.5 \times 10^{-7}$ \\
    O & $3.2 \times 10^{-4}$ & & Mg & $1.4 \times 10^{-7}$ \\
    \hline
  \end{tabular}
  \label{tab:abun}
\end{table}

\begin{figure*}
  \centering
  \includegraphics[width=0.31\textwidth]{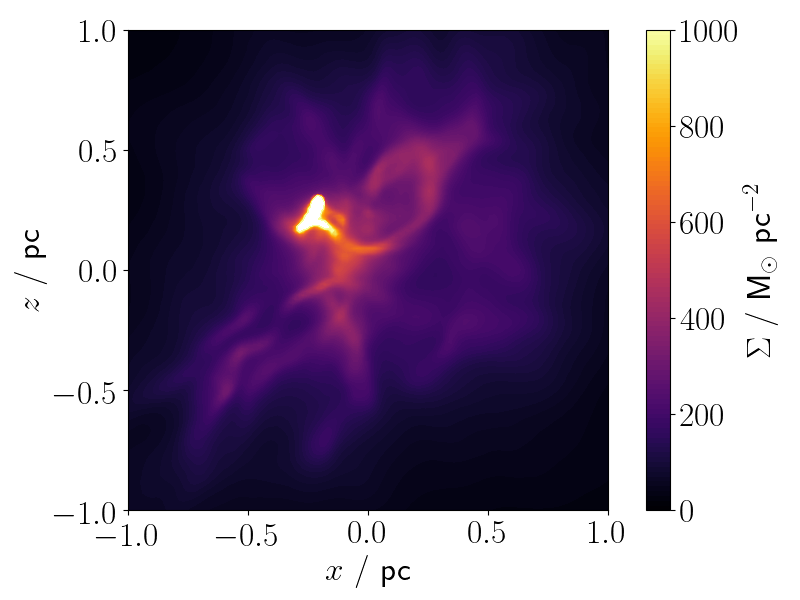}\quad
  \includegraphics[width=0.31\textwidth]{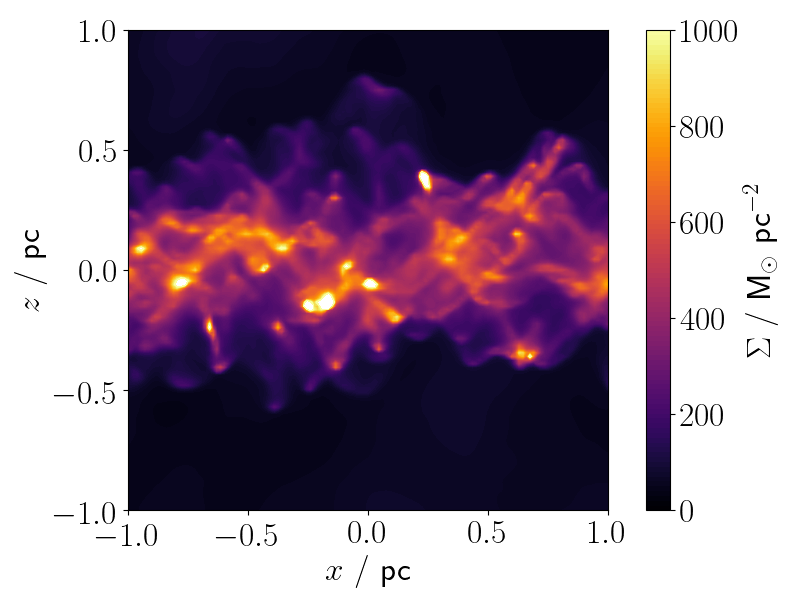}\quad
  \includegraphics[width=0.31\textwidth]{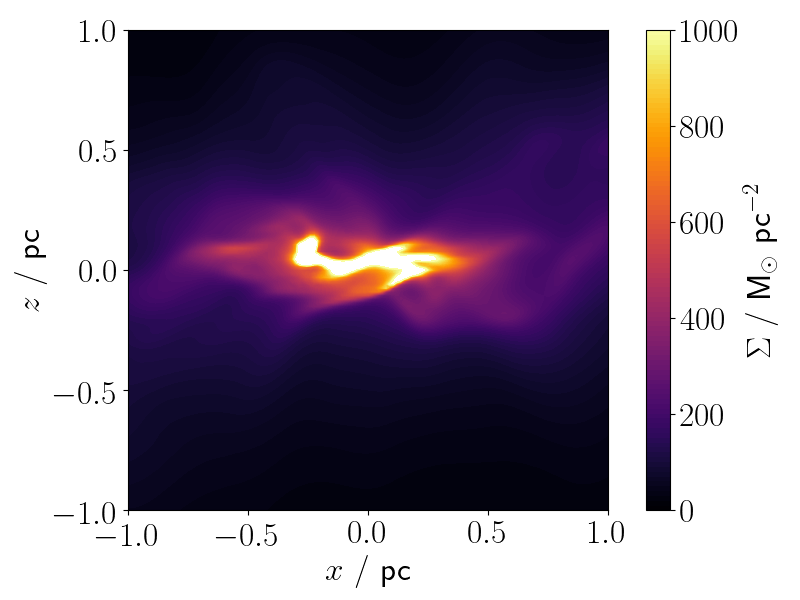}\\
  \includegraphics[width=0.31\textwidth]{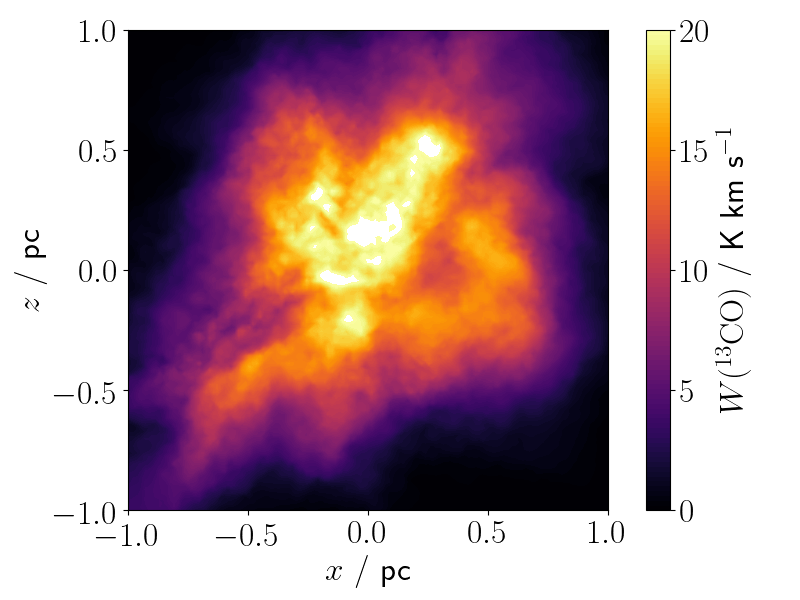}\quad
  \includegraphics[width=0.31\textwidth]{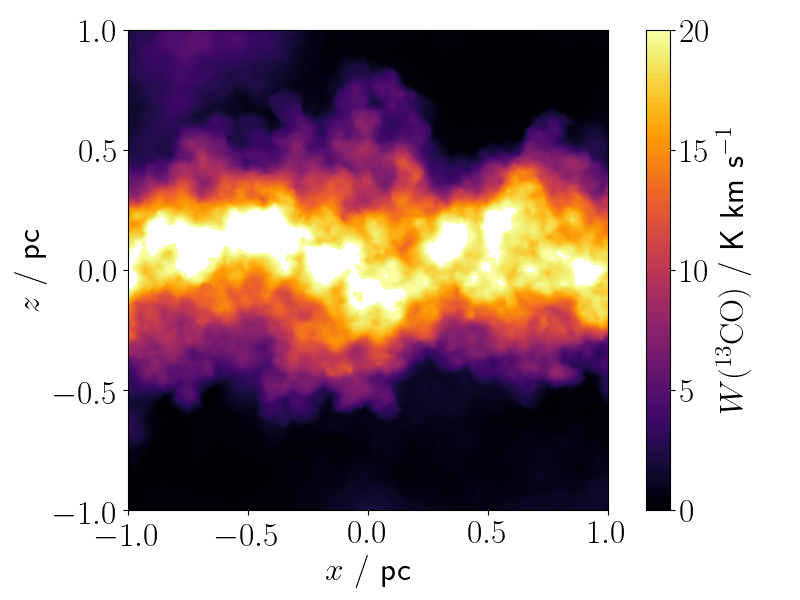}\quad
  \includegraphics[width=0.31\textwidth]{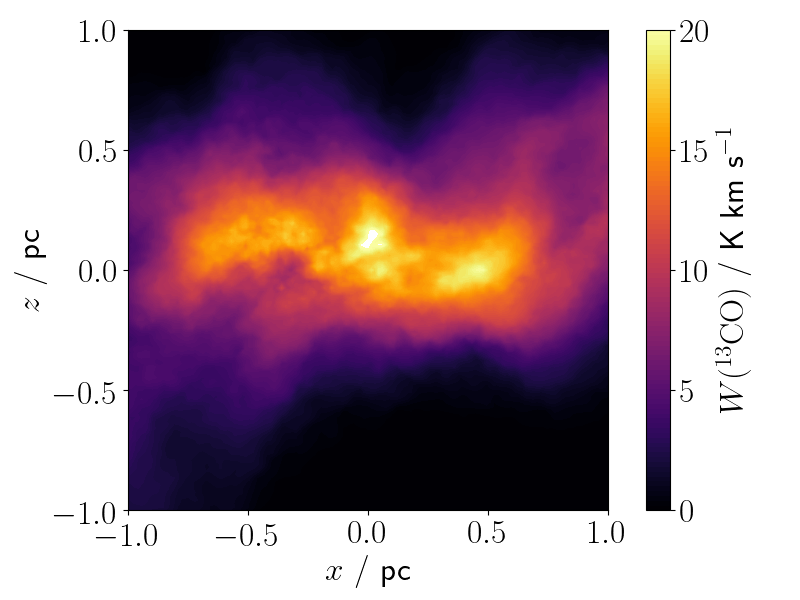}\\
  \includegraphics[width=0.31\textwidth]{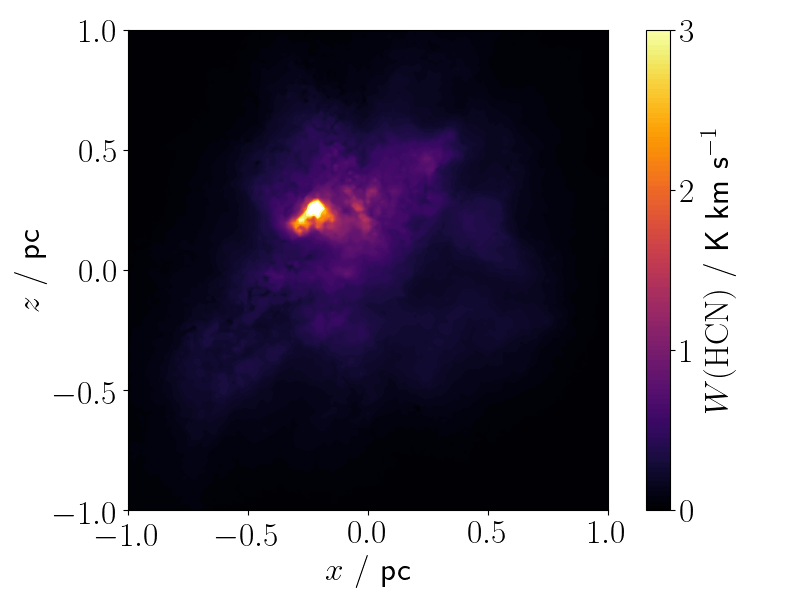}\quad
  \includegraphics[width=0.31\textwidth]{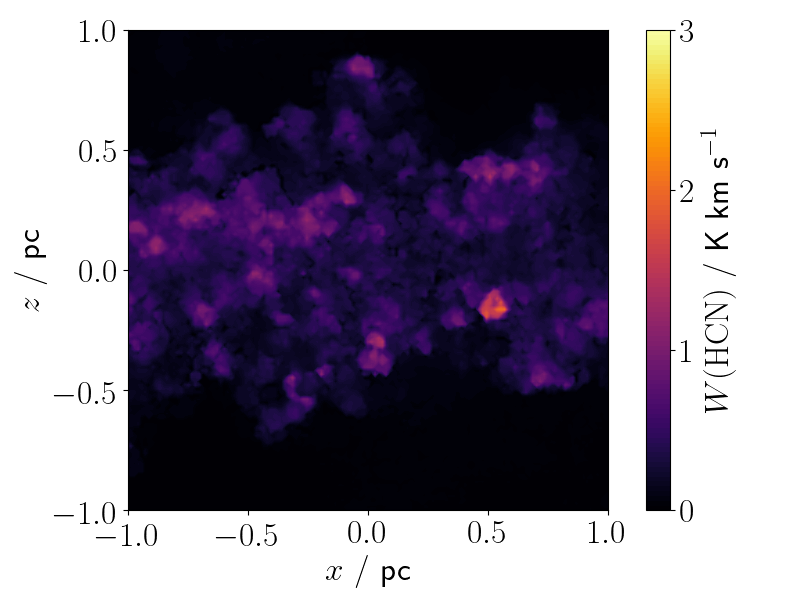}\quad
  \includegraphics[width=0.31\textwidth]{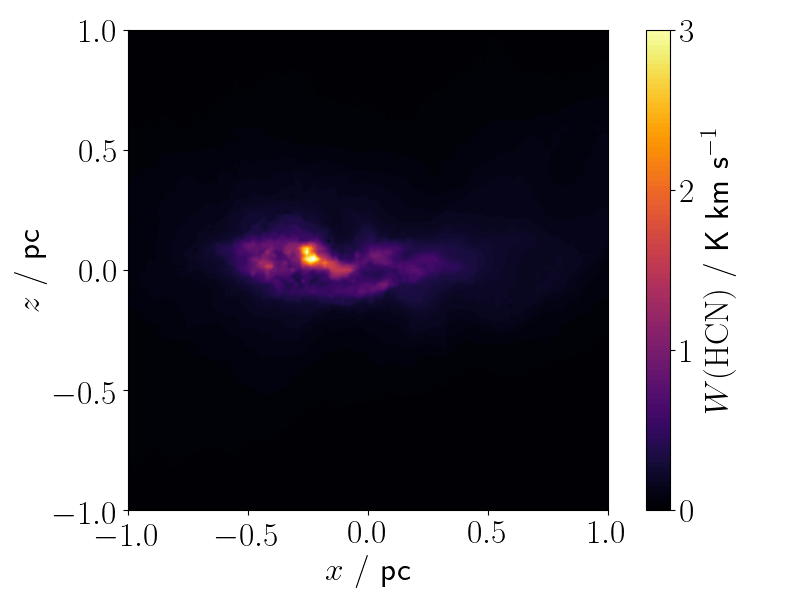}\\
  \caption{Column density (top row) and integrated intensities of the $^{13}$CO $J=1-0$ (middle row) and HCN $J=1-0$ (bottom row) lines, seen in the $x-z$ plane for the TURB (left column), COLL (middle column) and MAG (right column) models.}
  \label{fig:img}
\end{figure*}

\begin{figure*}
  \centering
  \includegraphics[width=0.31\textwidth]{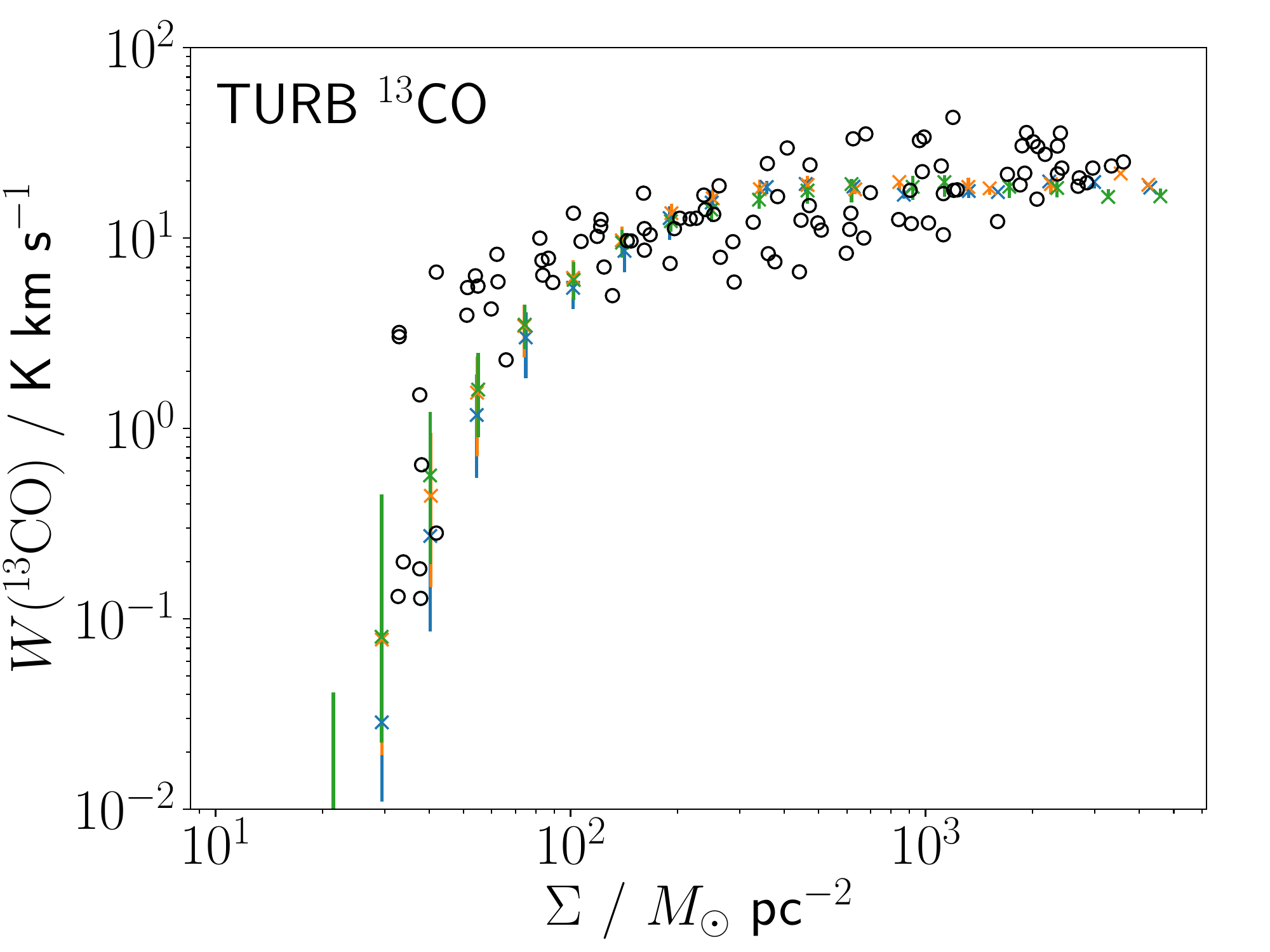}\quad
  \includegraphics[width=0.31\textwidth]{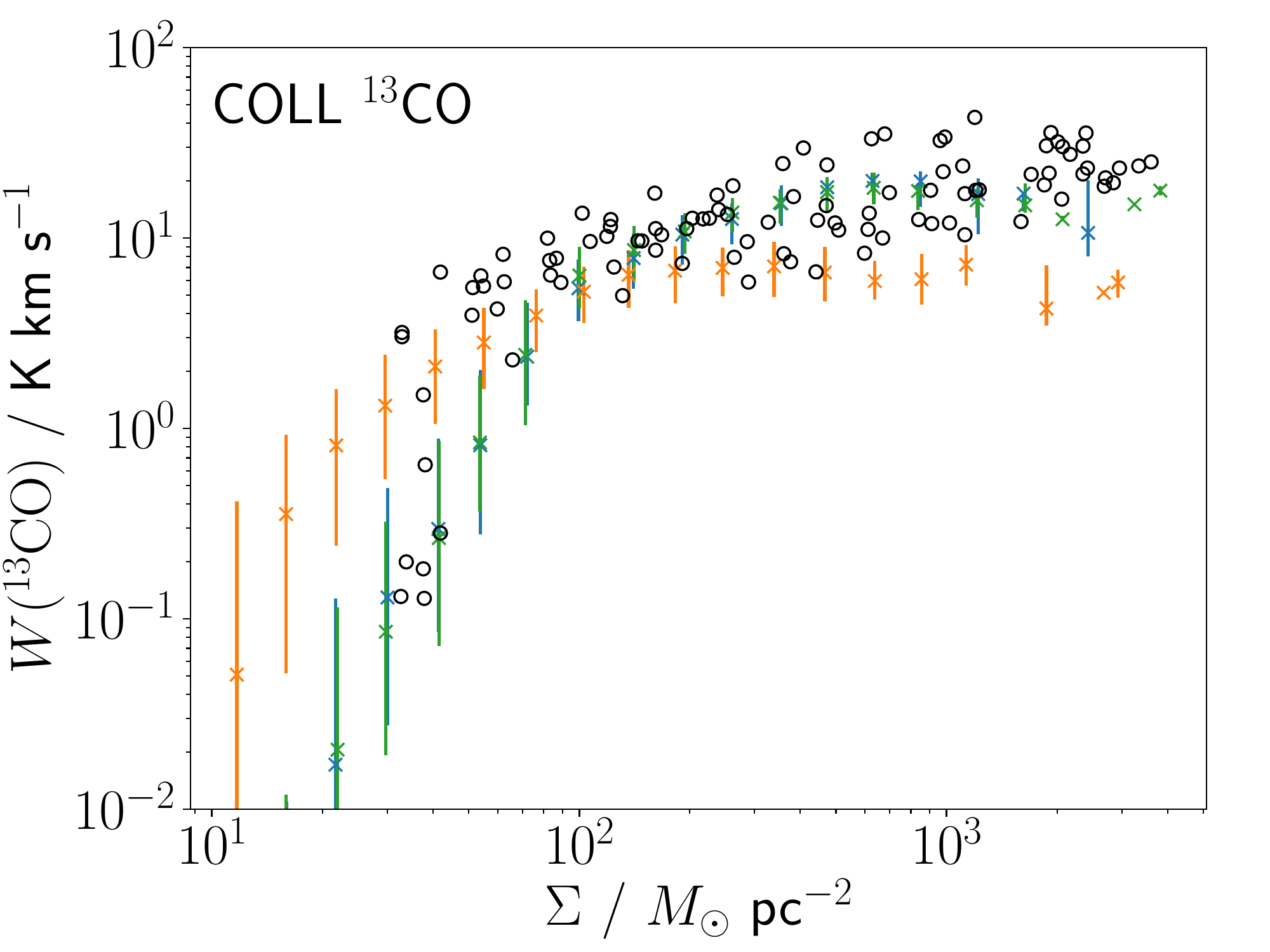}\quad
  \includegraphics[width=0.31\textwidth]{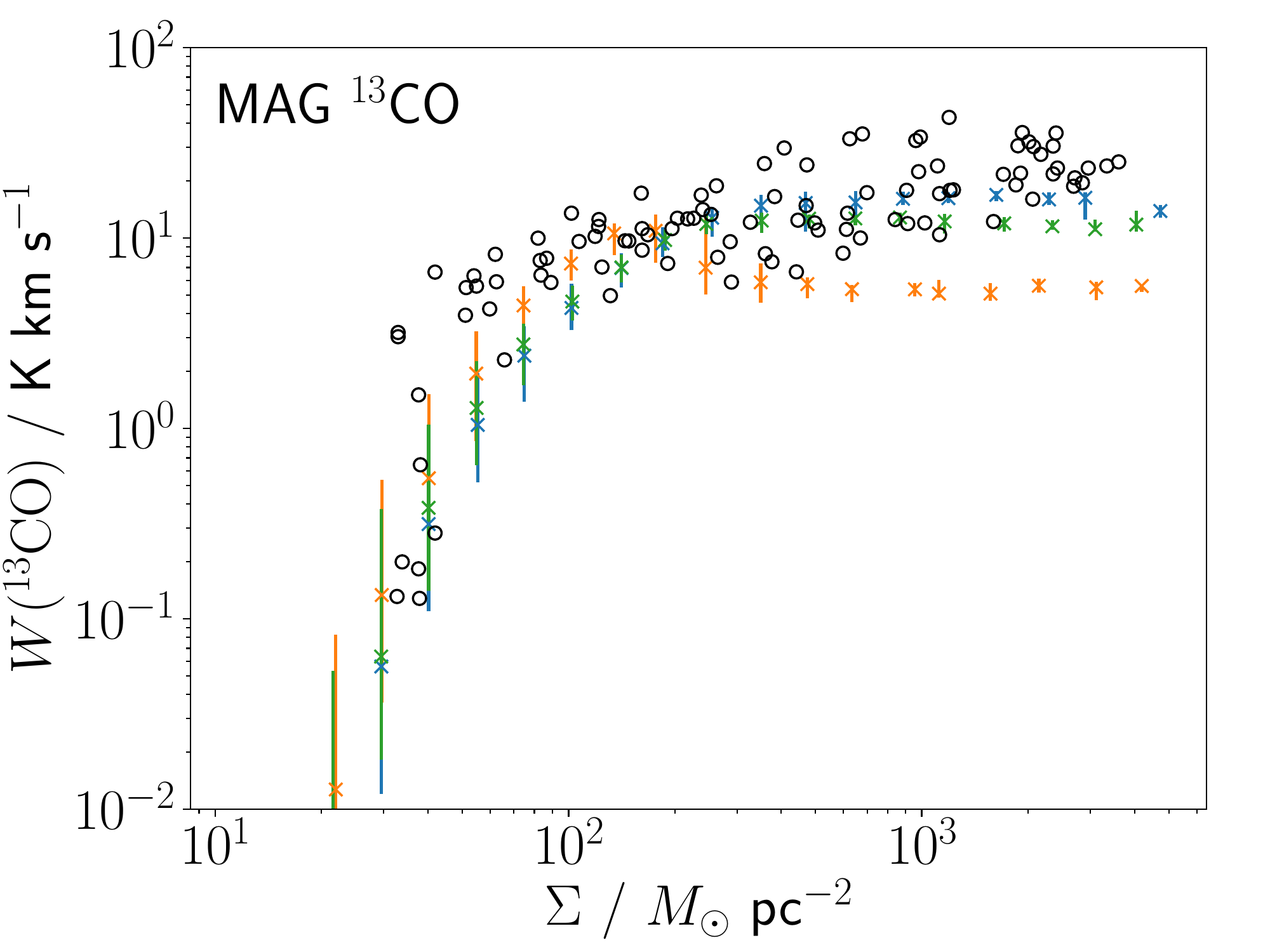}\\
  \includegraphics[width=0.31\textwidth]{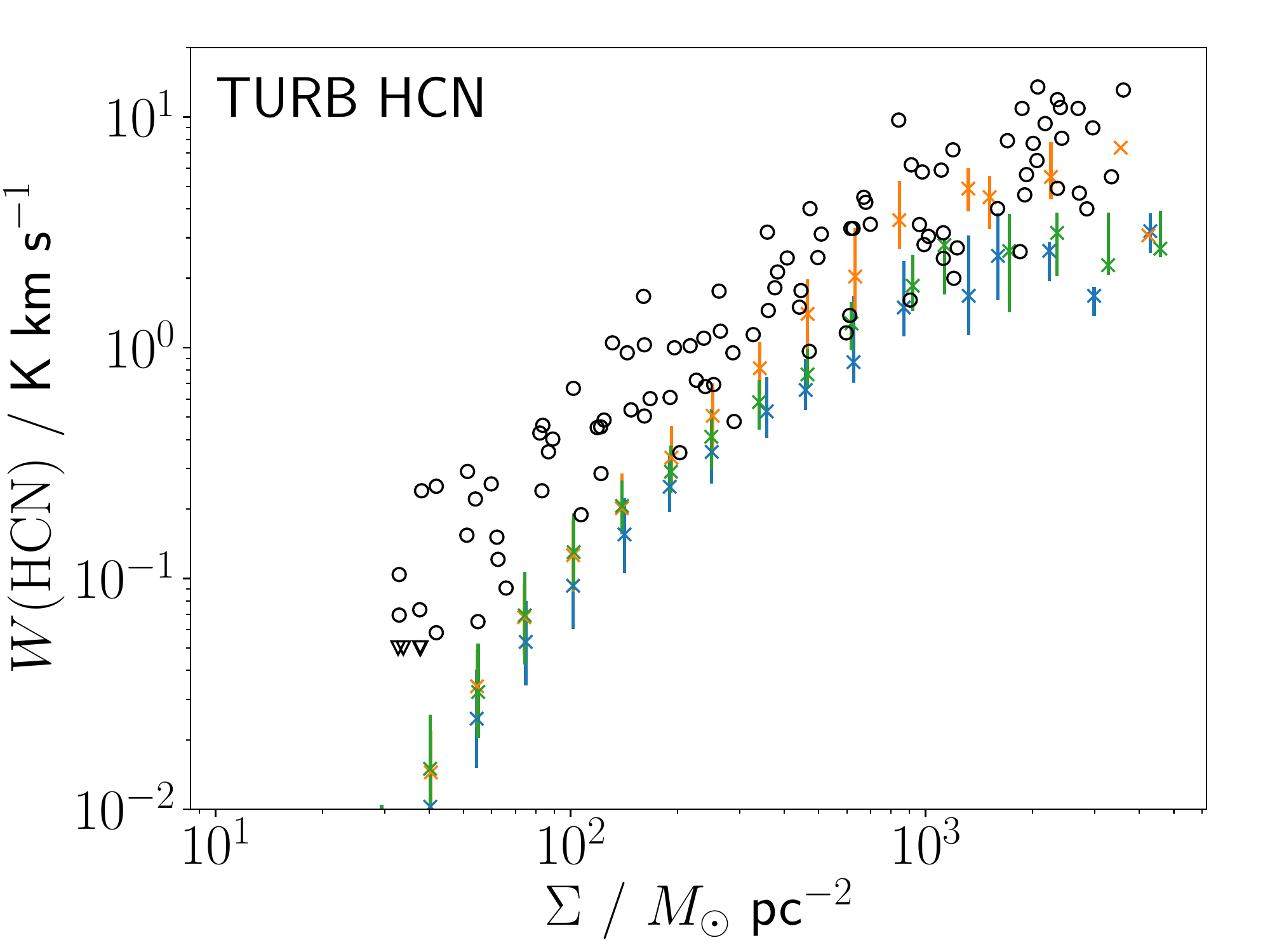}\quad
  \includegraphics[width=0.31\textwidth]{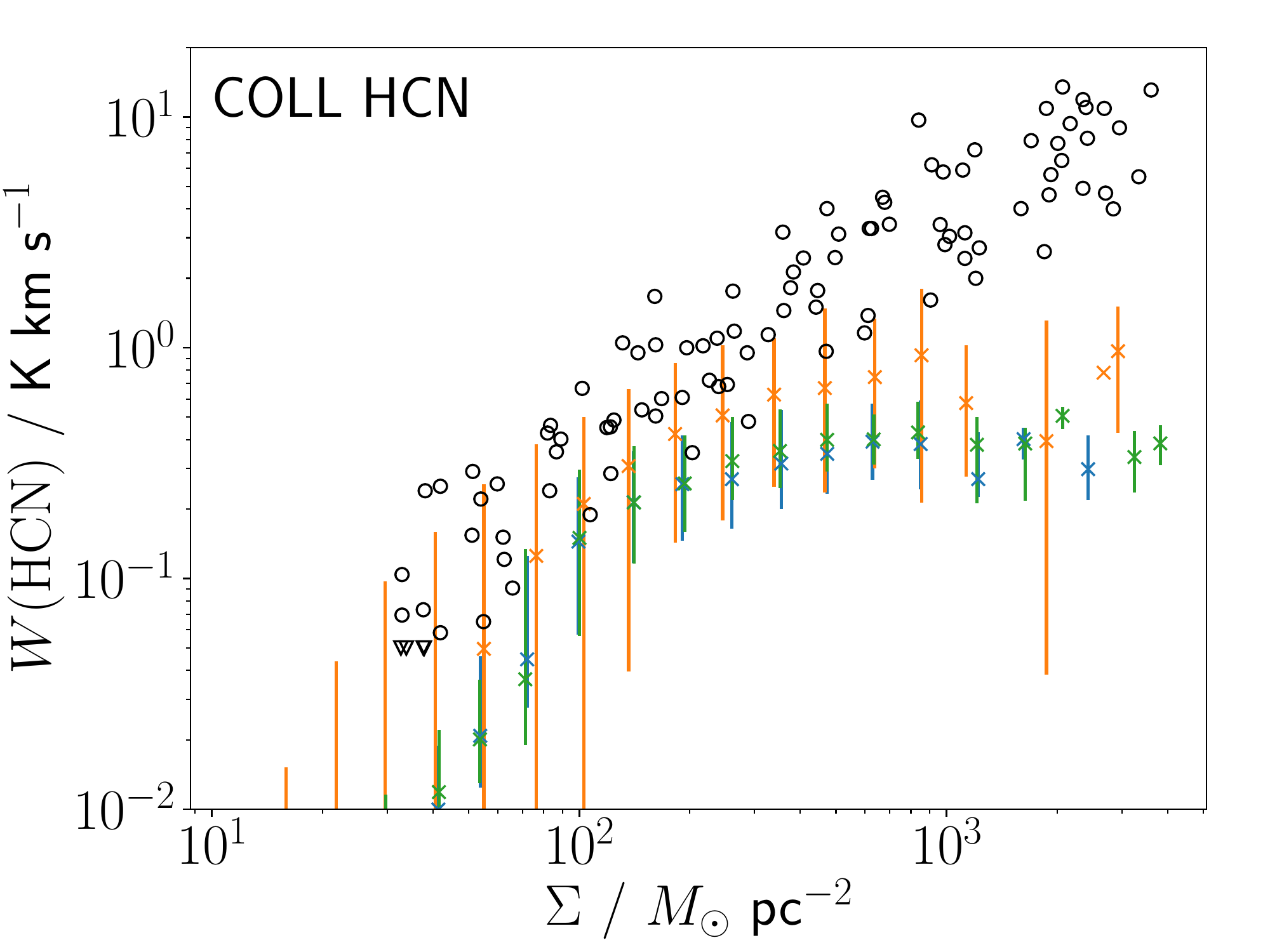}\quad
  \includegraphics[width=0.31\textwidth]{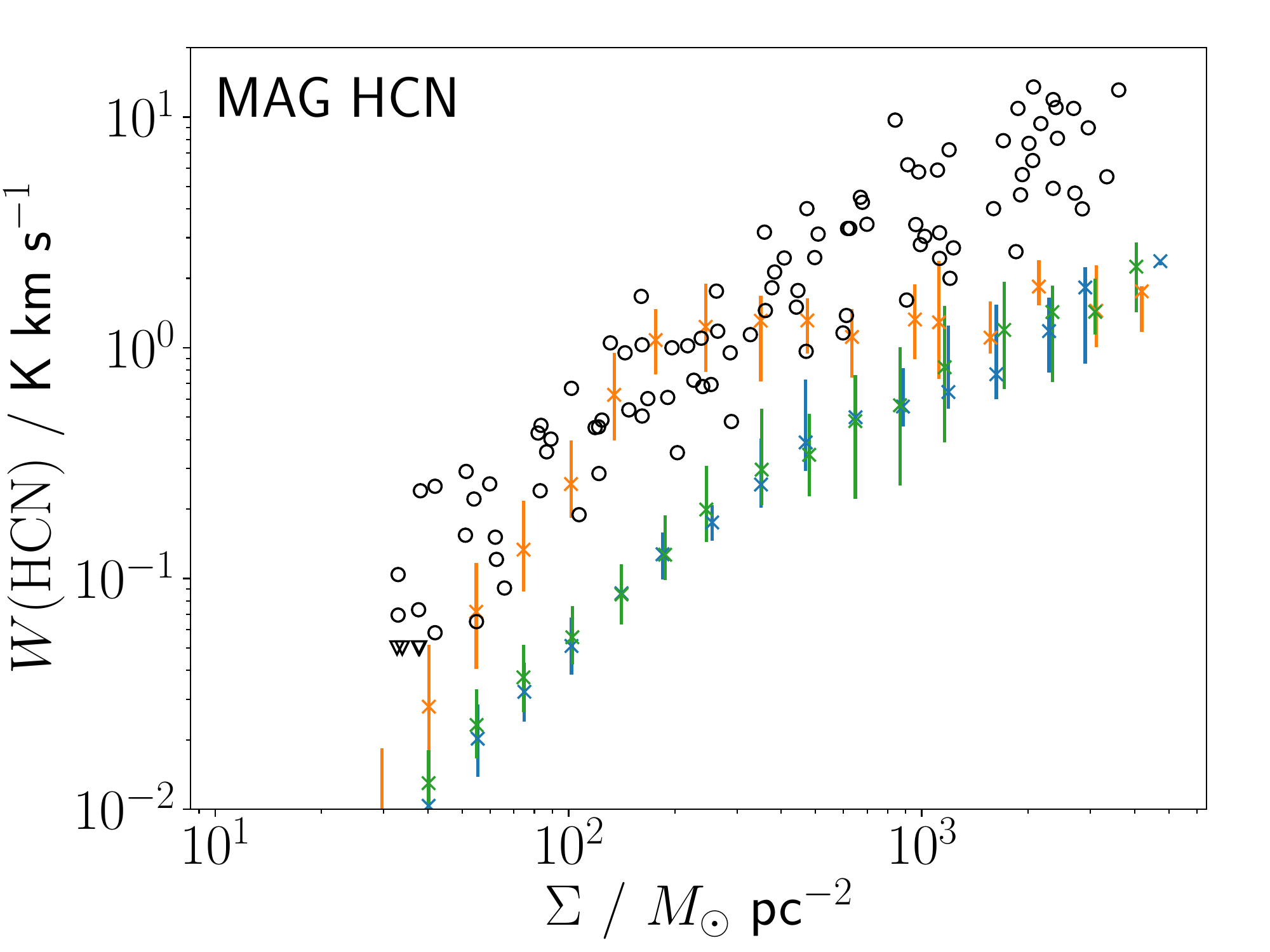}\\
  \includegraphics[width=0.31\textwidth]{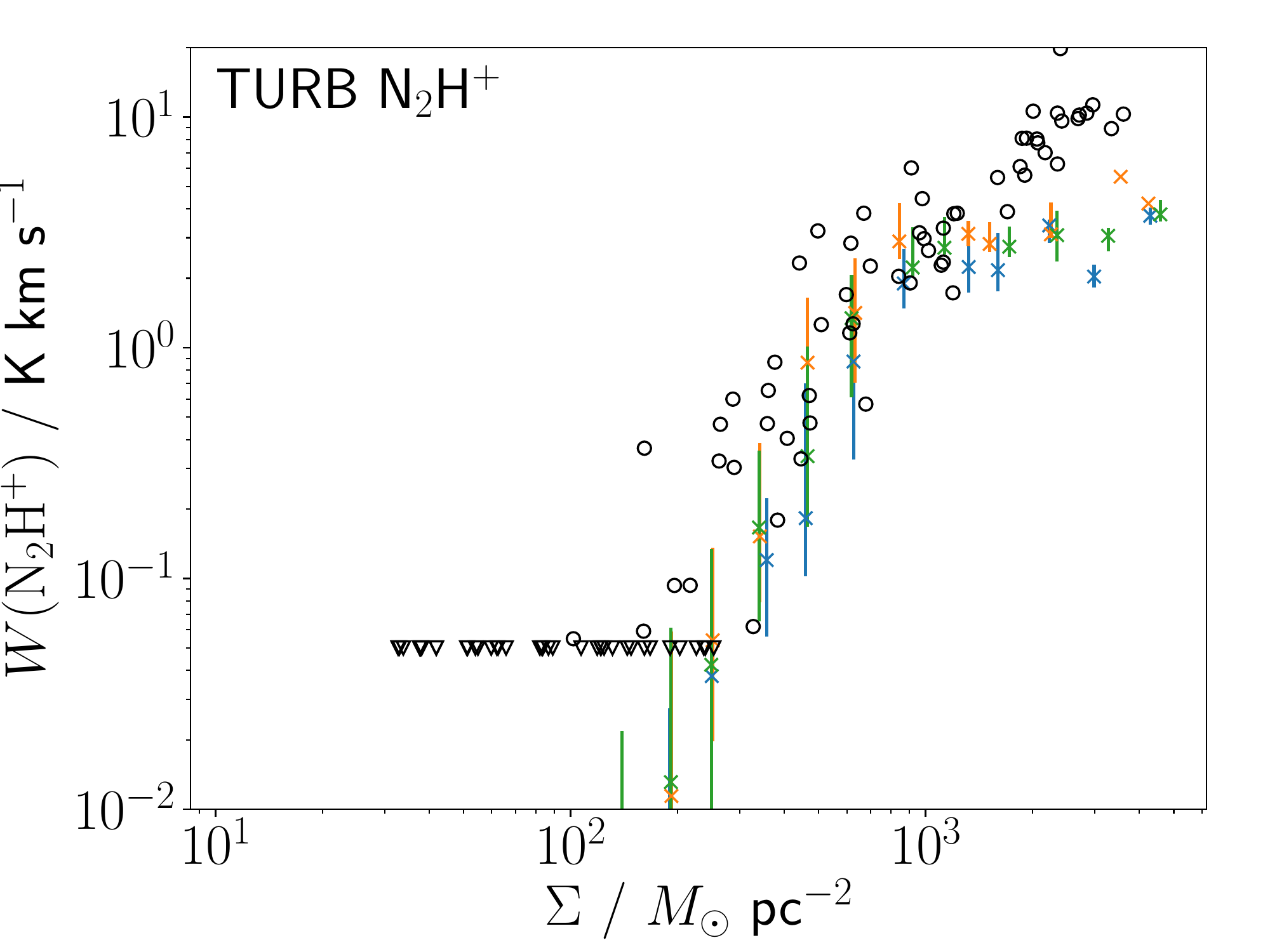}\quad
  \includegraphics[width=0.31\textwidth]{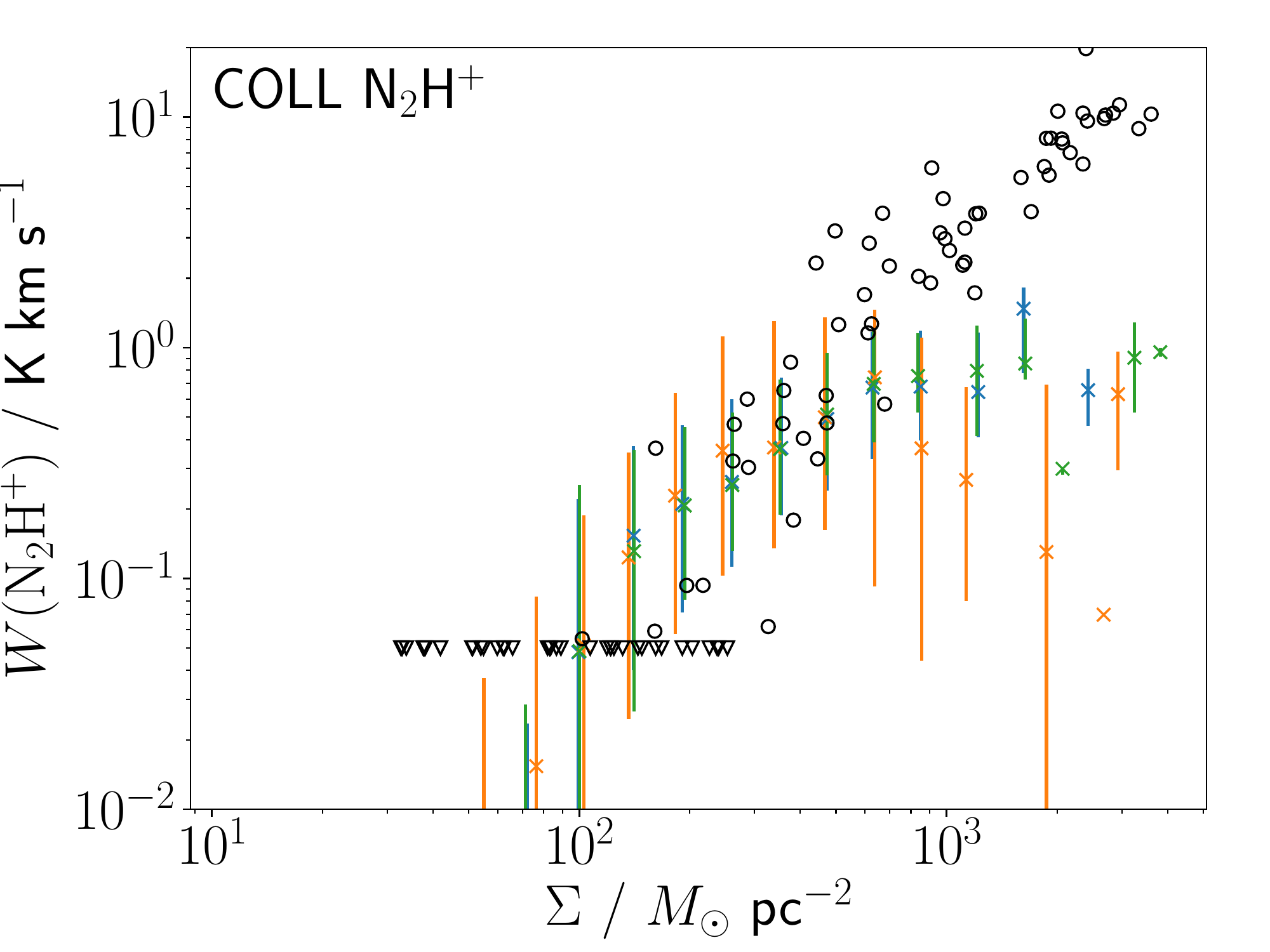}\quad
  \includegraphics[width=0.31\textwidth]{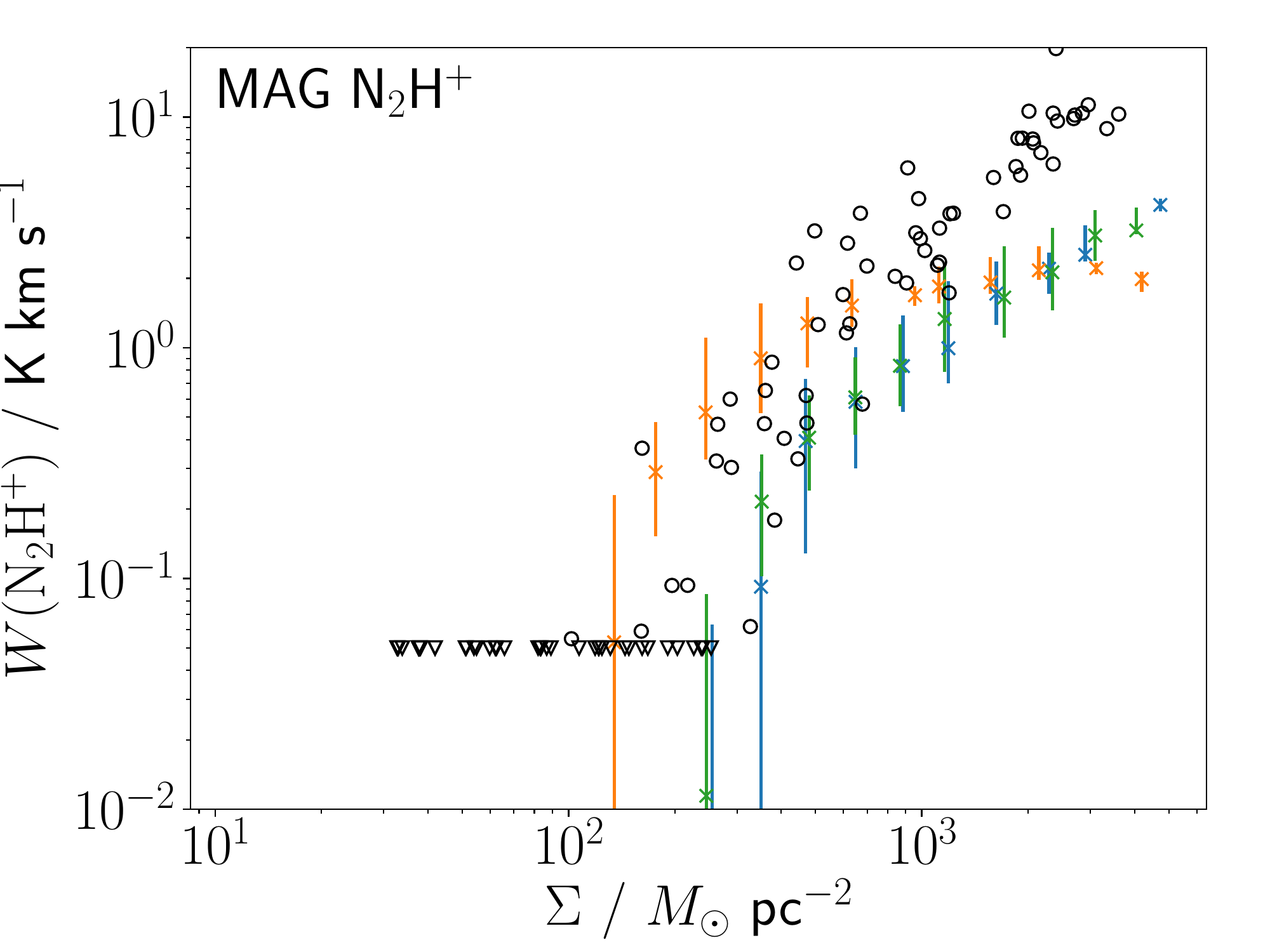}\\
  \caption{Integrated line intensity versus column density for the TURB (left), COLL (centre) and MAG (right) models, seen in the $x-y$ (orange), $x-z$ (blue) and $y-z$ (green) planes. {\it Top:} $^{13}$CO $J=1-0$; {\it Middle:} HCN $J=1-0$; {\it Bottom:} N$_2$H$^+$ $J=1-0$. Median values are shown as crosses, with the 16th and 84th percentiles as error bars. Observations from \citet{tafalla2021} are shown as circles, with those below the assumed detection limit of $0.05 \kel$ shown as triangles.}
  \label{fig:scale}
\end{figure*}

\begin{figure}
  \centering
  \includegraphics[width=\columnwidth]{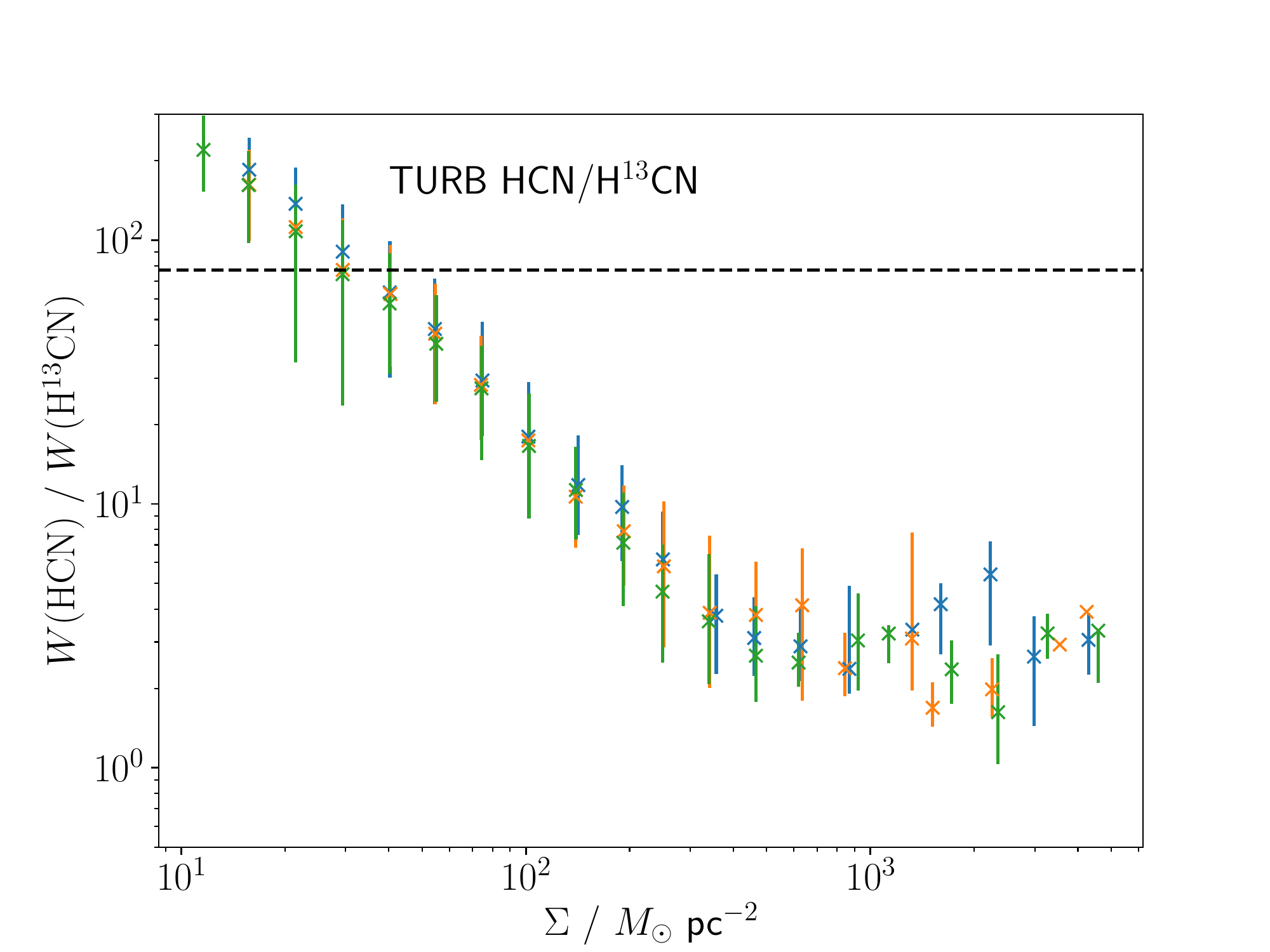}
  \caption{Ratio of HCN to H$^{13}$CN $J=1-0$ integrated intensity versus column density for the TURB model, seen in the $x-y$ (orange), $x-z$ (blue) and $y-z$ (green) planes. Median values are shown as crosses, with the 16th and 84th percentiles as error bars. The dashed line shows the adopted abundance ratio of $77$, which should be replicated by the intensity ratio for optically thin emission.}
  \label{fig:hcnratio}
\end{figure}

\begin{figure*}
  \centering
  \includegraphics[width=0.31\textwidth]{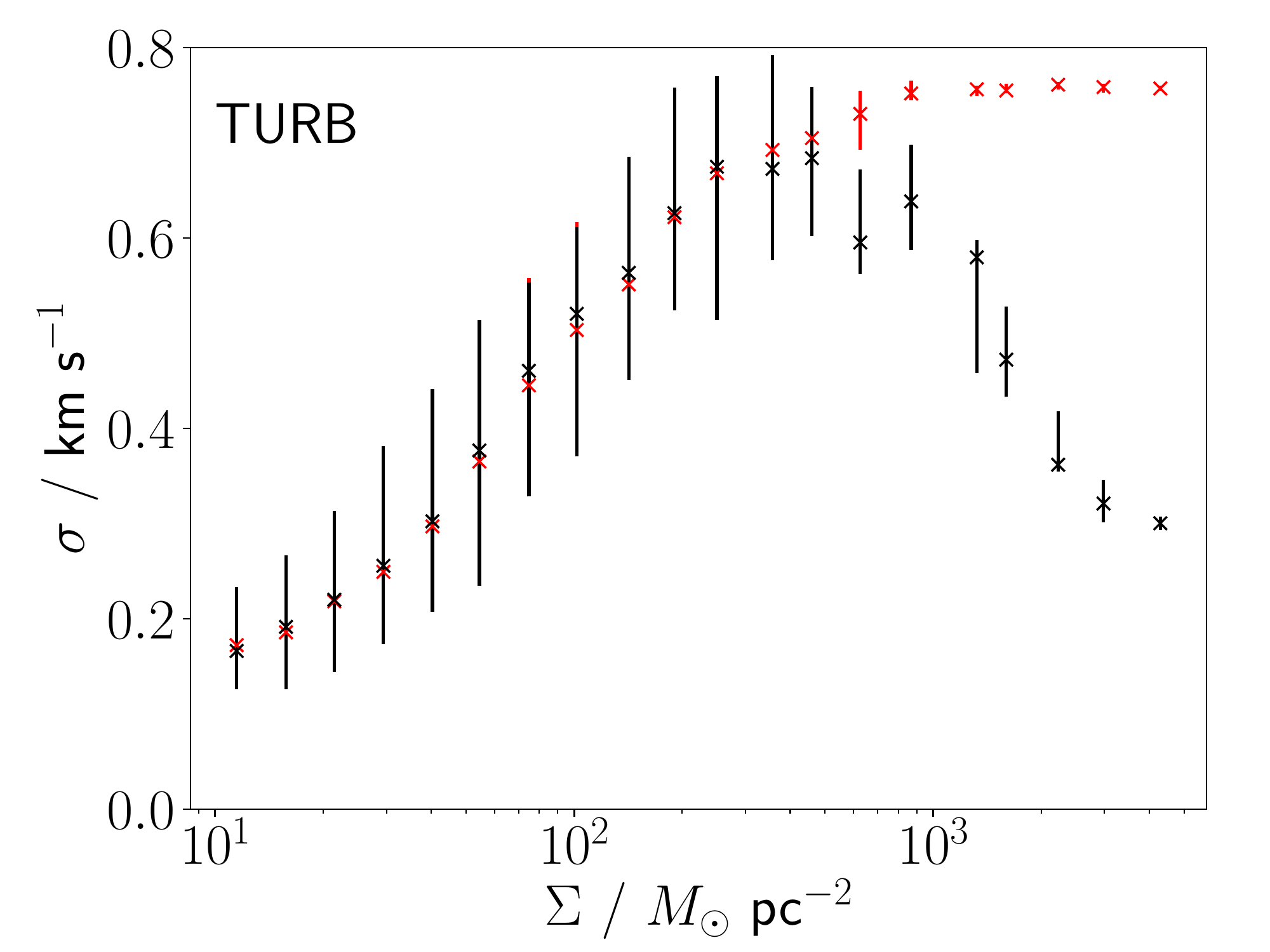}\quad
  \includegraphics[width=0.31\textwidth]{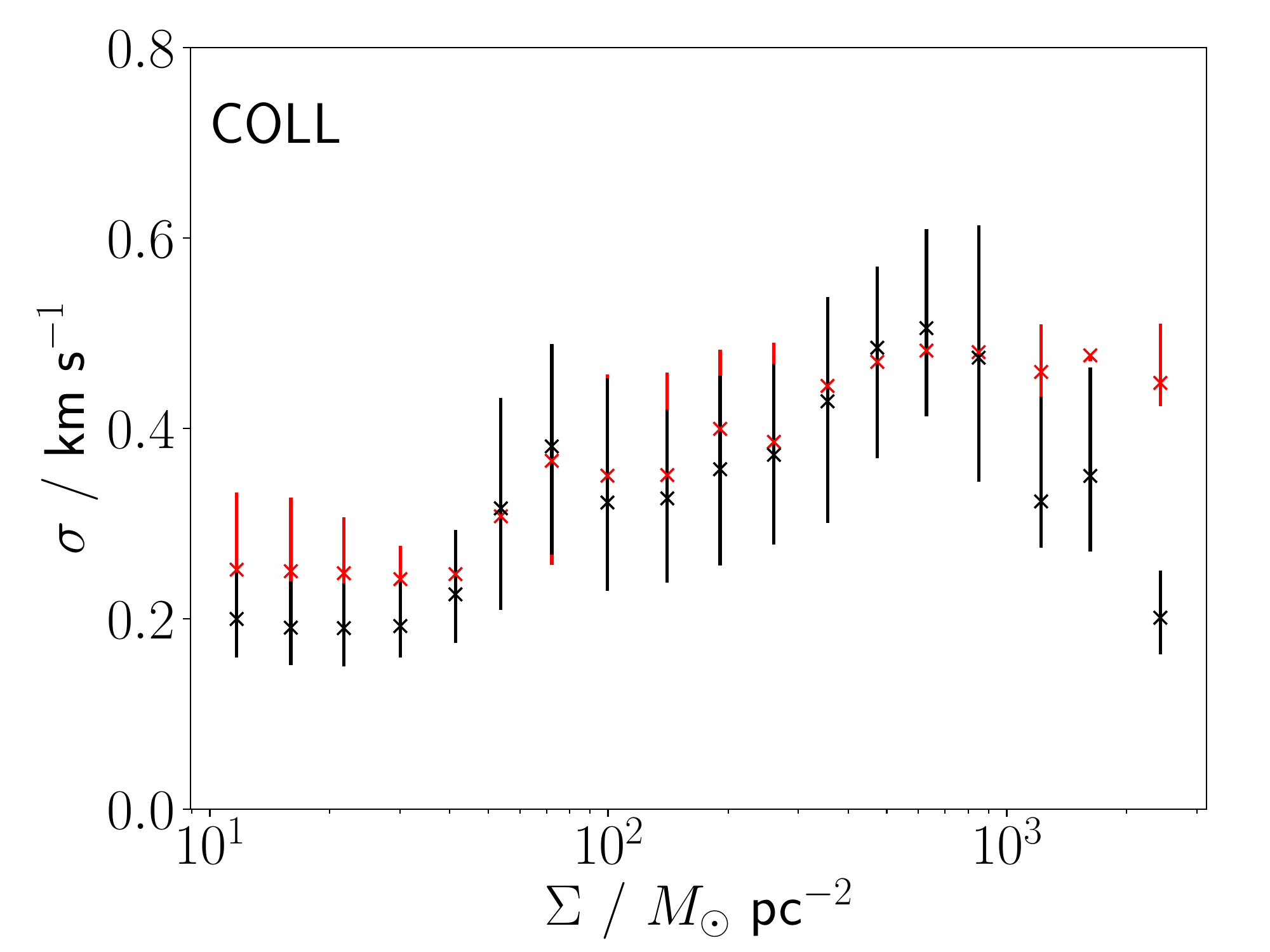}\quad
  \includegraphics[width=0.31\textwidth]{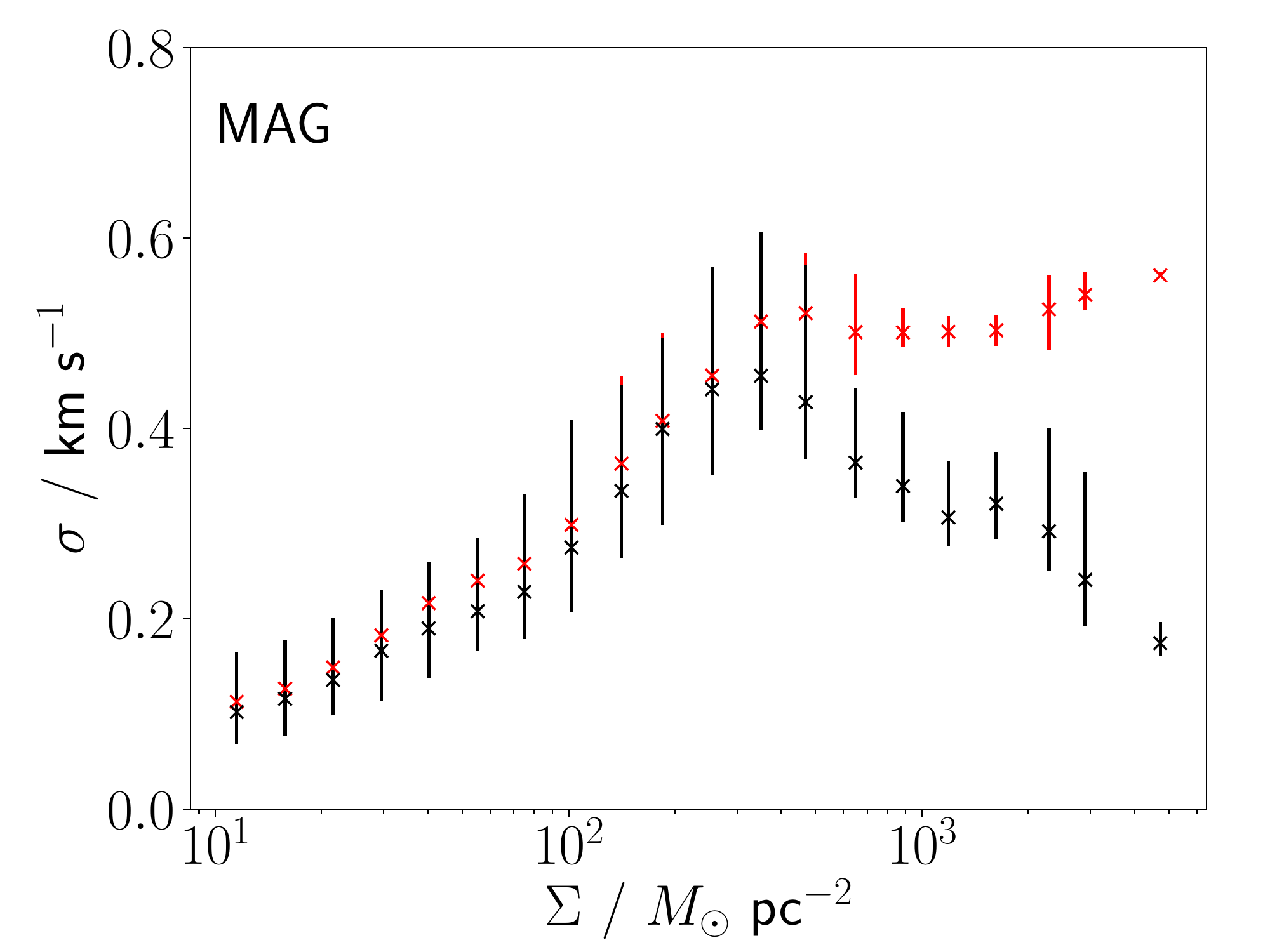}\\
  \caption{Volume-weighted (red) and mass-weighted (black) line-of-sight velocity dispersions versus column density for the TURB (left), COLL (centre) and MAG (right) models, seen in the $x-z$ plane. Median values are shown as crosses, with the 16th and 84th percentiles as error bars.}
  \label{fig:sigv}
\end{figure*}

\begin{figure}
  \centering
  \includegraphics[width=\columnwidth]{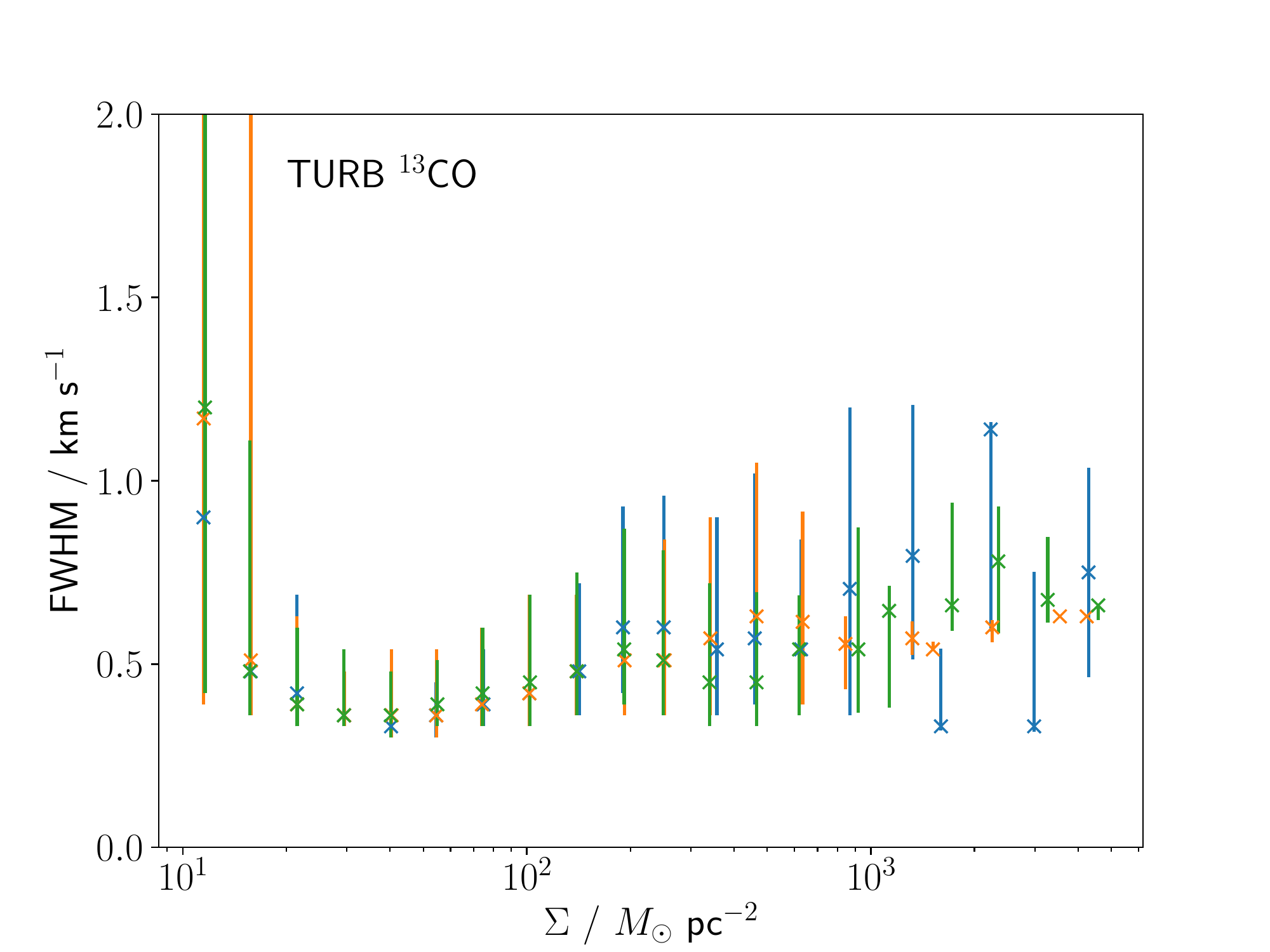}
  \caption{$^{13}$CO $J=1-0$ line full-width at half-maximum versus column density for the TURB model, seen in the $x-y$ (orange), $x-z$ (blue) and $y-z$ (green) planes. Median values are shown as crosses, with the 16th and 84th percentiles as error bars.}
  \label{fig:fwhm}
\end{figure}

\begin{figure*}
  \centering
  \includegraphics[width=0.31\textwidth]{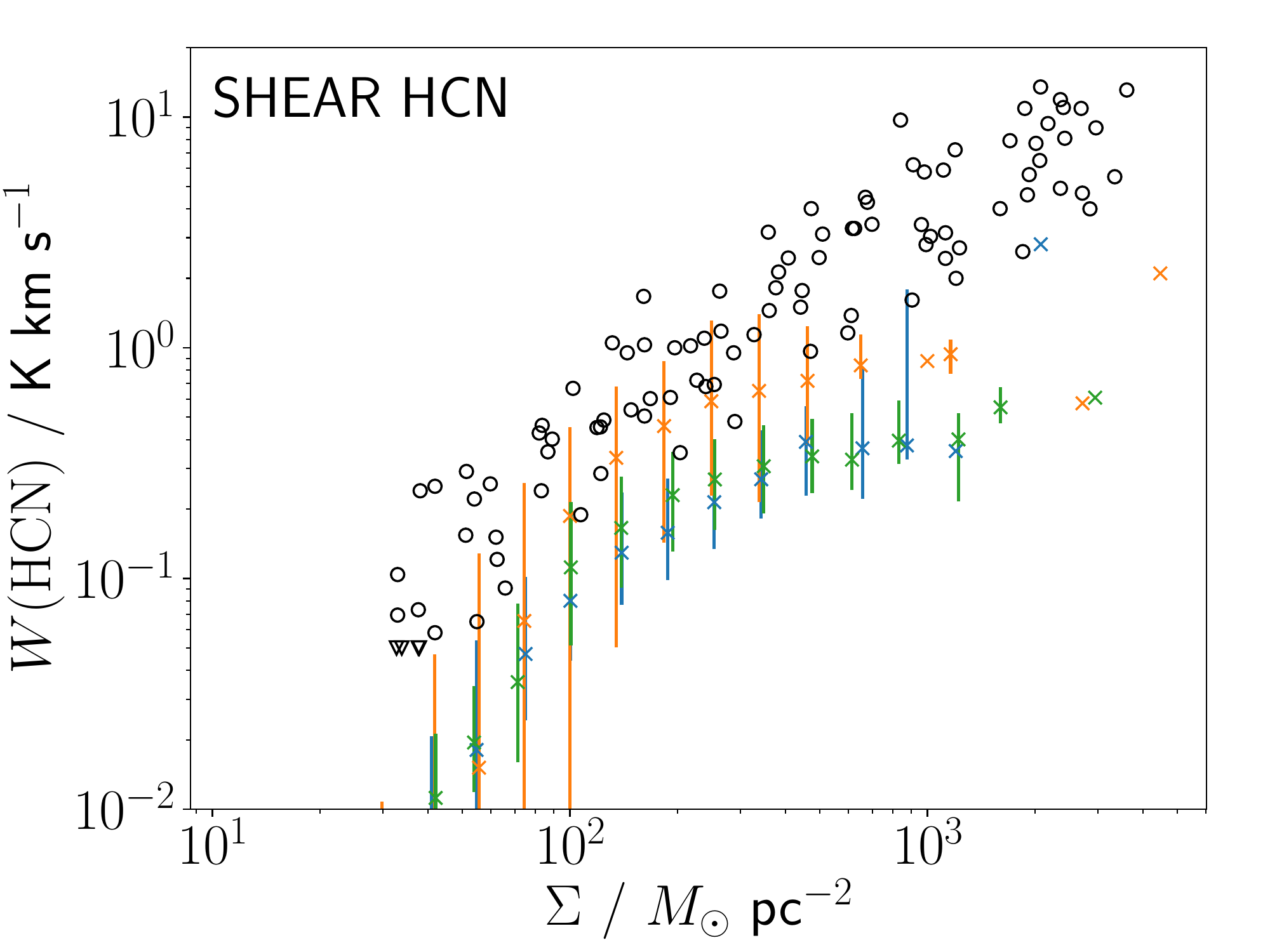}\quad
  \includegraphics[width=0.31\textwidth]{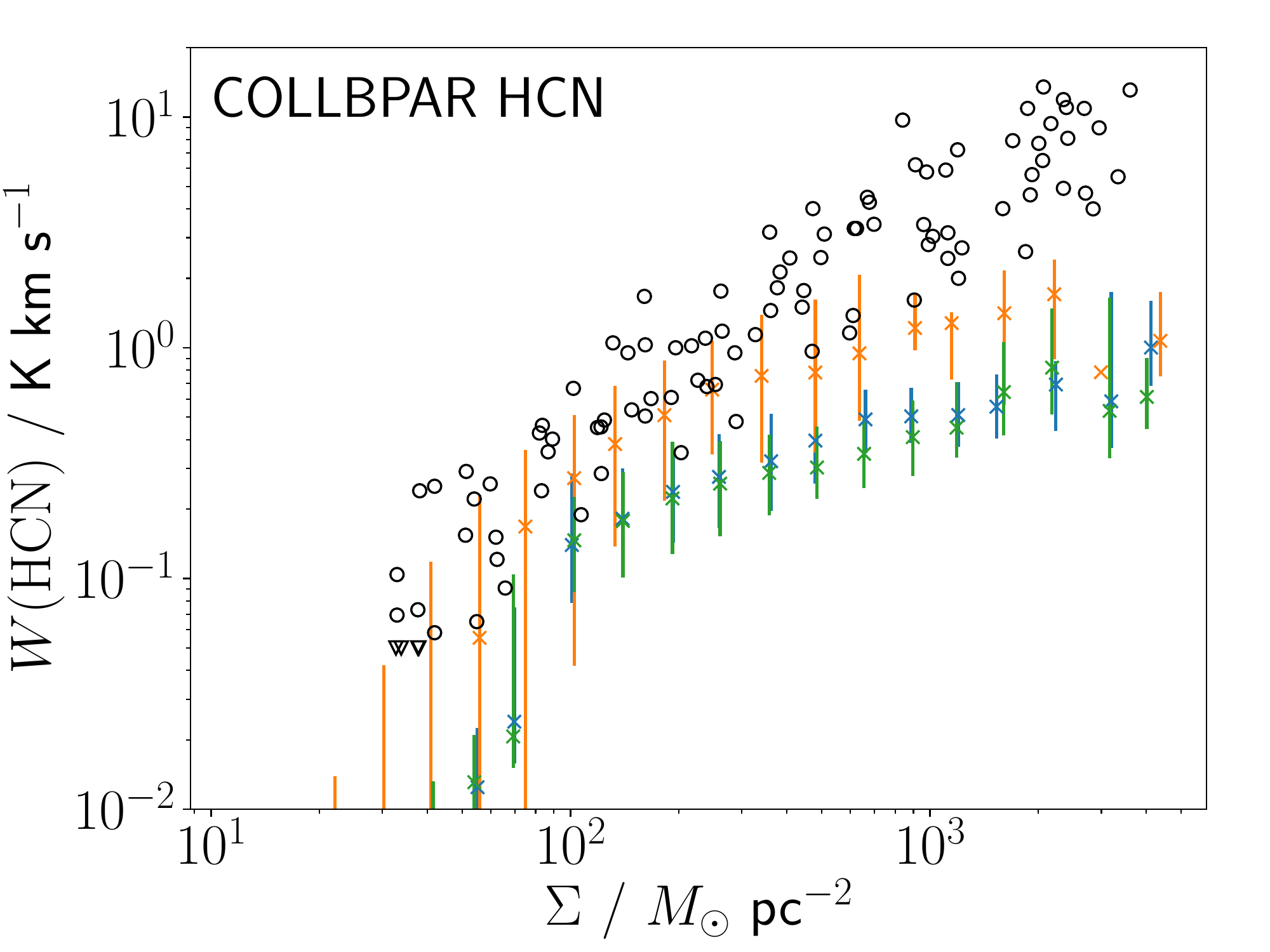}\quad
  \includegraphics[width=0.31\textwidth]{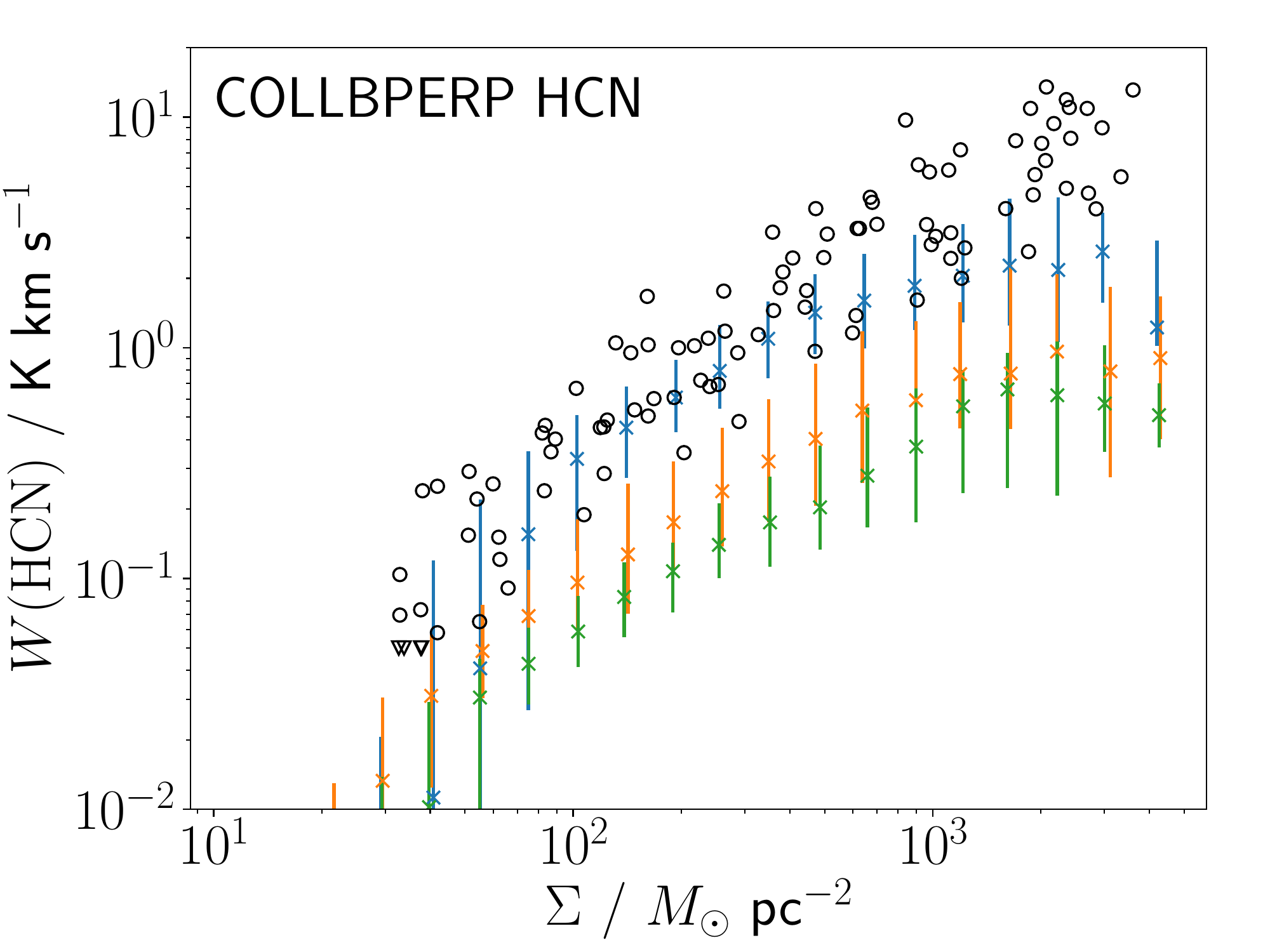}\\
  \caption{Integrated HCN $J=1-0$ line intensity versus column density for the SHEAR (left), COLLBPAR (centre) and COLLBPERP (right) models, seen in the $x-y$ (orange), $x-z$ (blue) and $y-z$ (green) planes. Median values are shown as crosses, with the 16th and 84th percentiles as error bars. Observations from \citet{tafalla2021} are shown as circles, with those below the assumed detection limit of $0.05 \kel$ shown as triangles.}
  \label{fig:collscale}
\end{figure*}

\begin{figure*}
  \centering
  \includegraphics[width=0.31\textwidth]{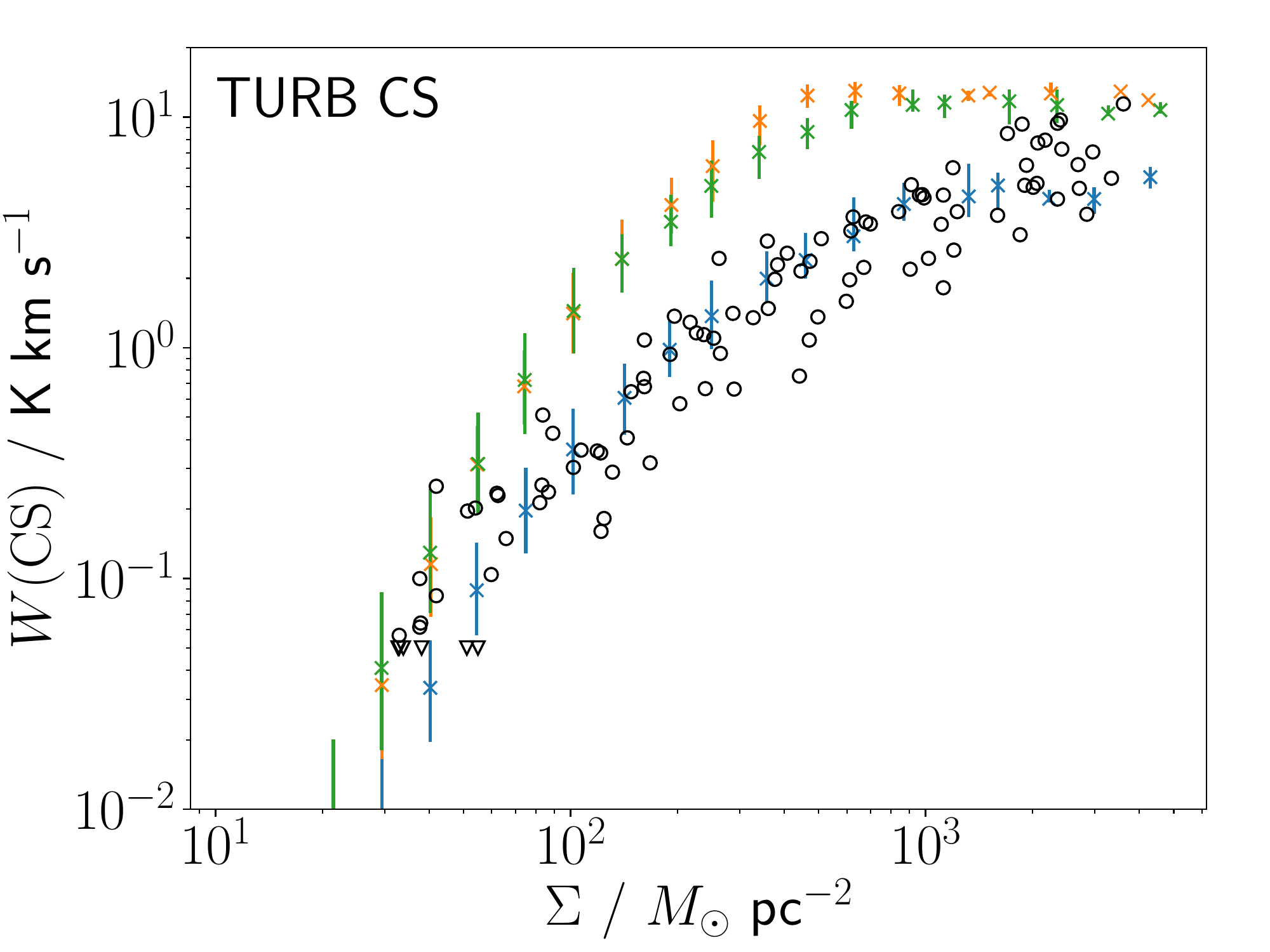}\quad
  \includegraphics[width=0.31\textwidth]{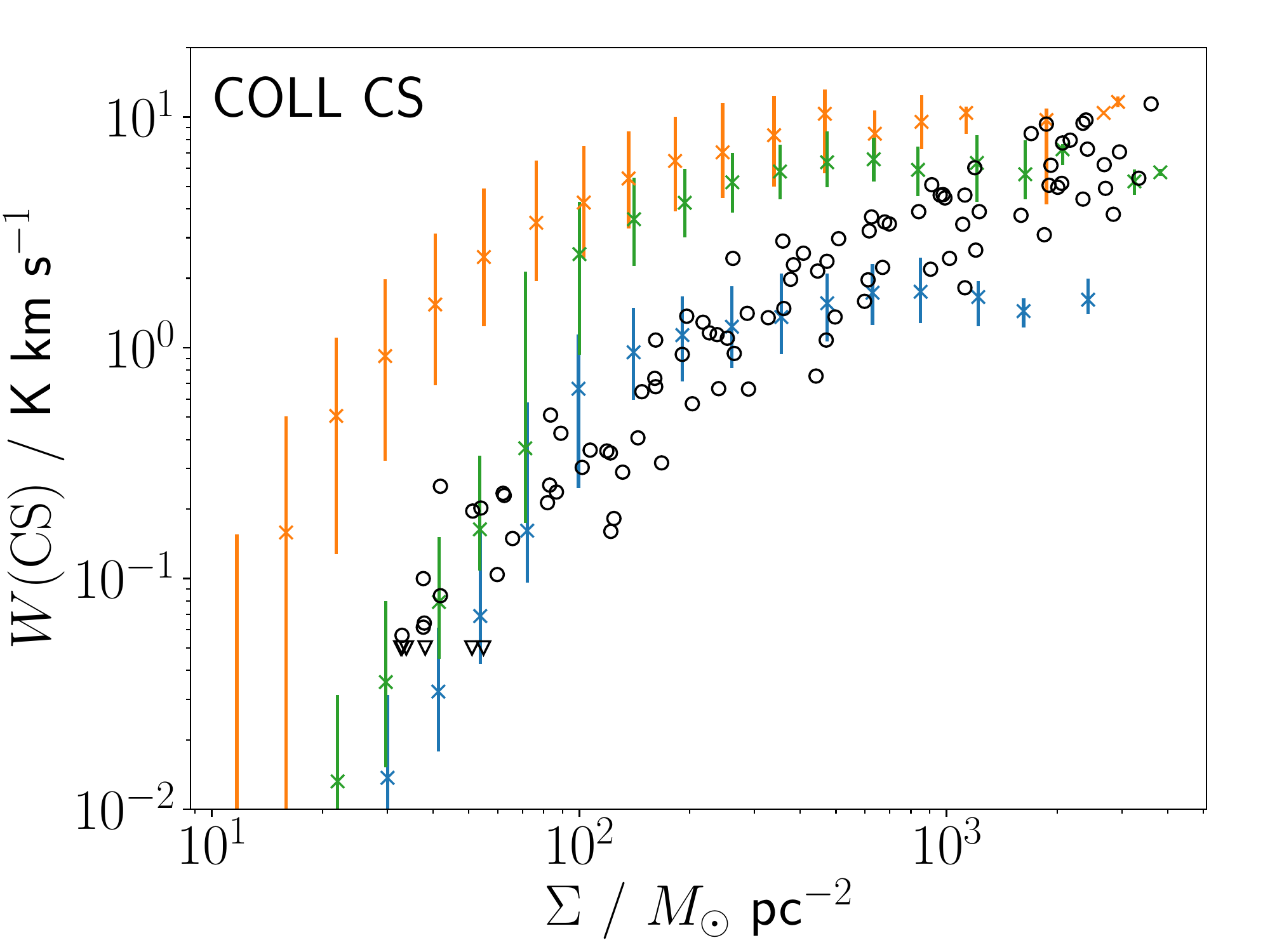}\quad
  \includegraphics[width=0.31\textwidth]{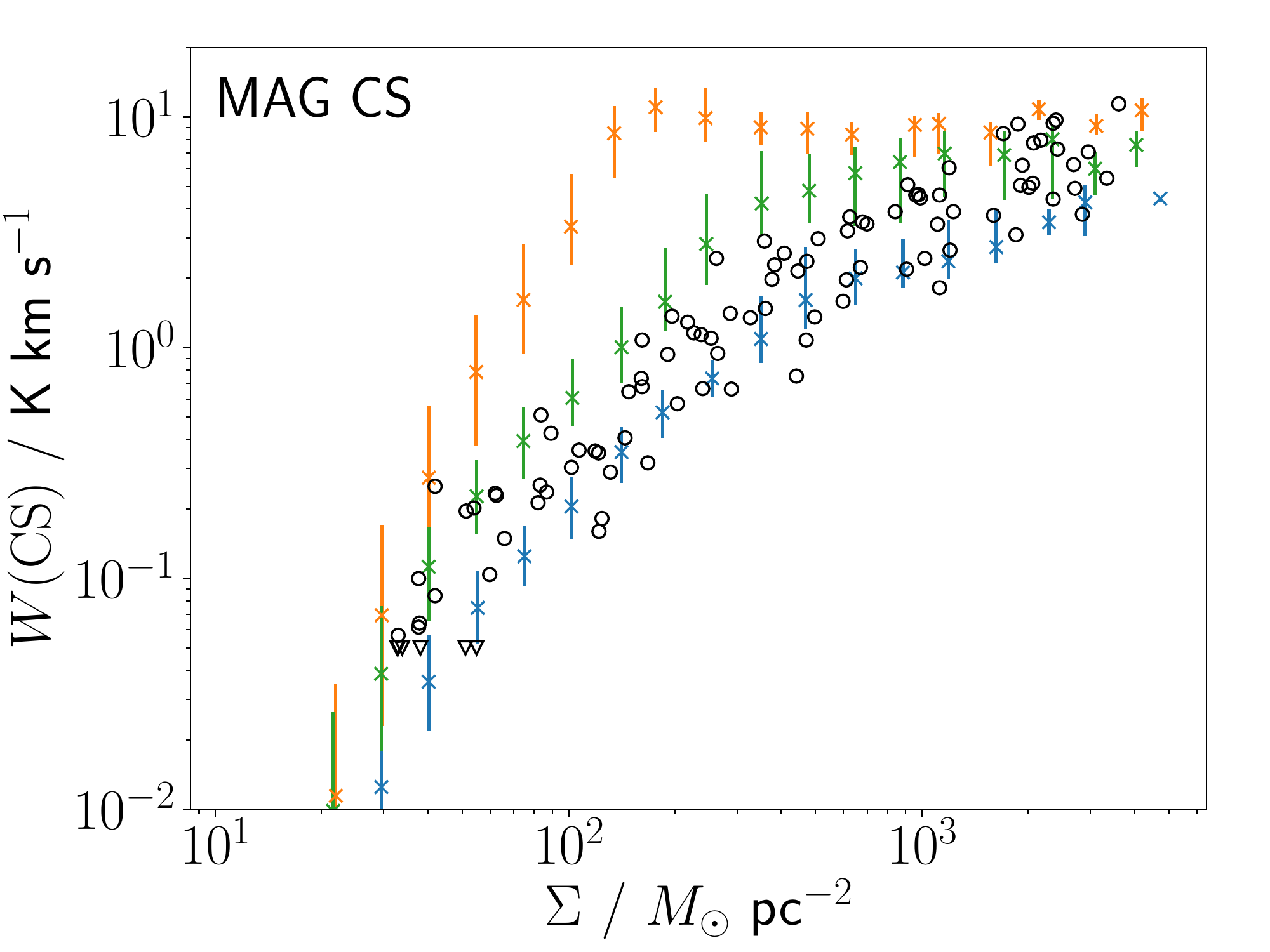}\\
  \includegraphics[width=0.31\textwidth]{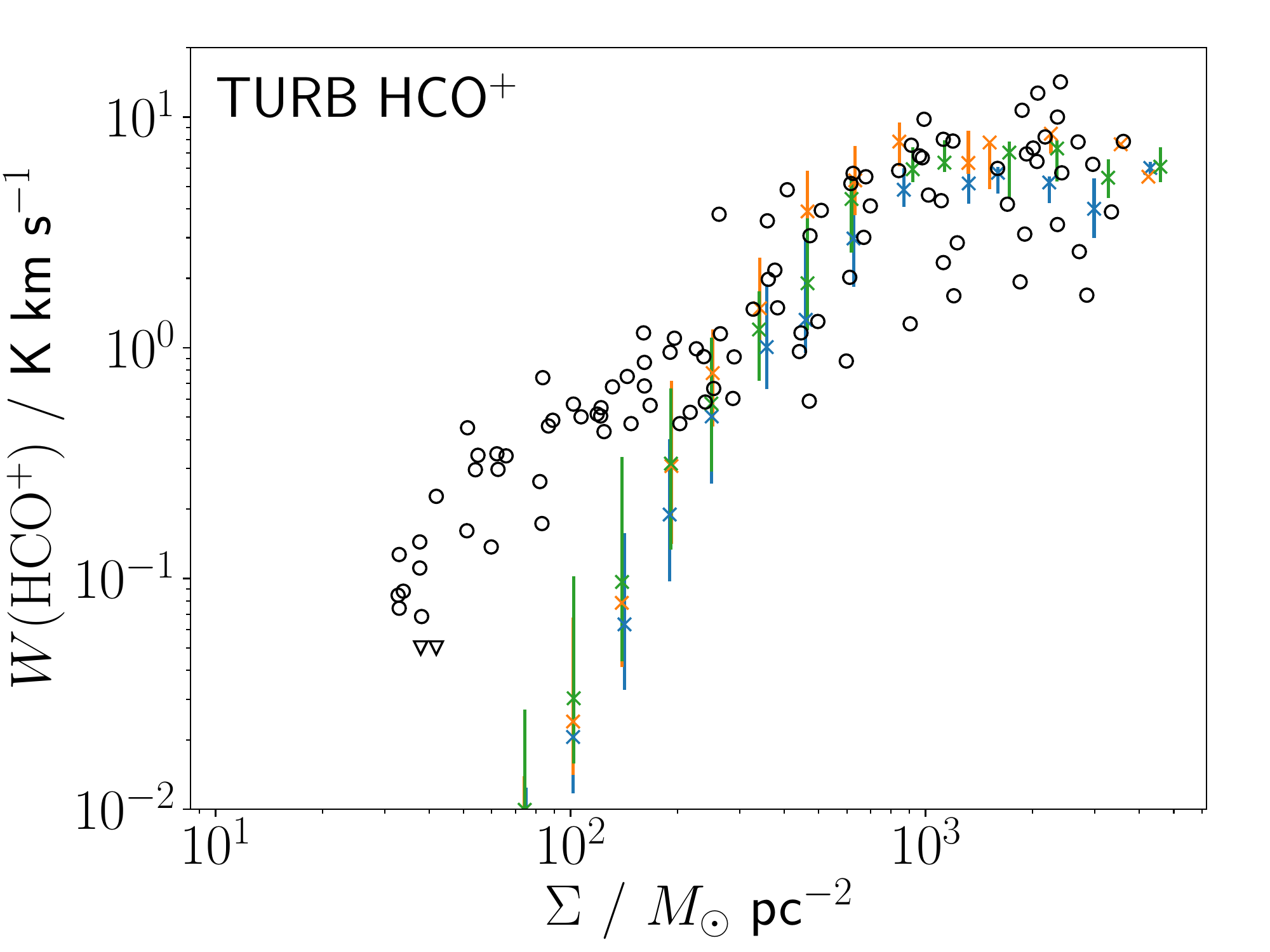}\quad
  \includegraphics[width=0.31\textwidth]{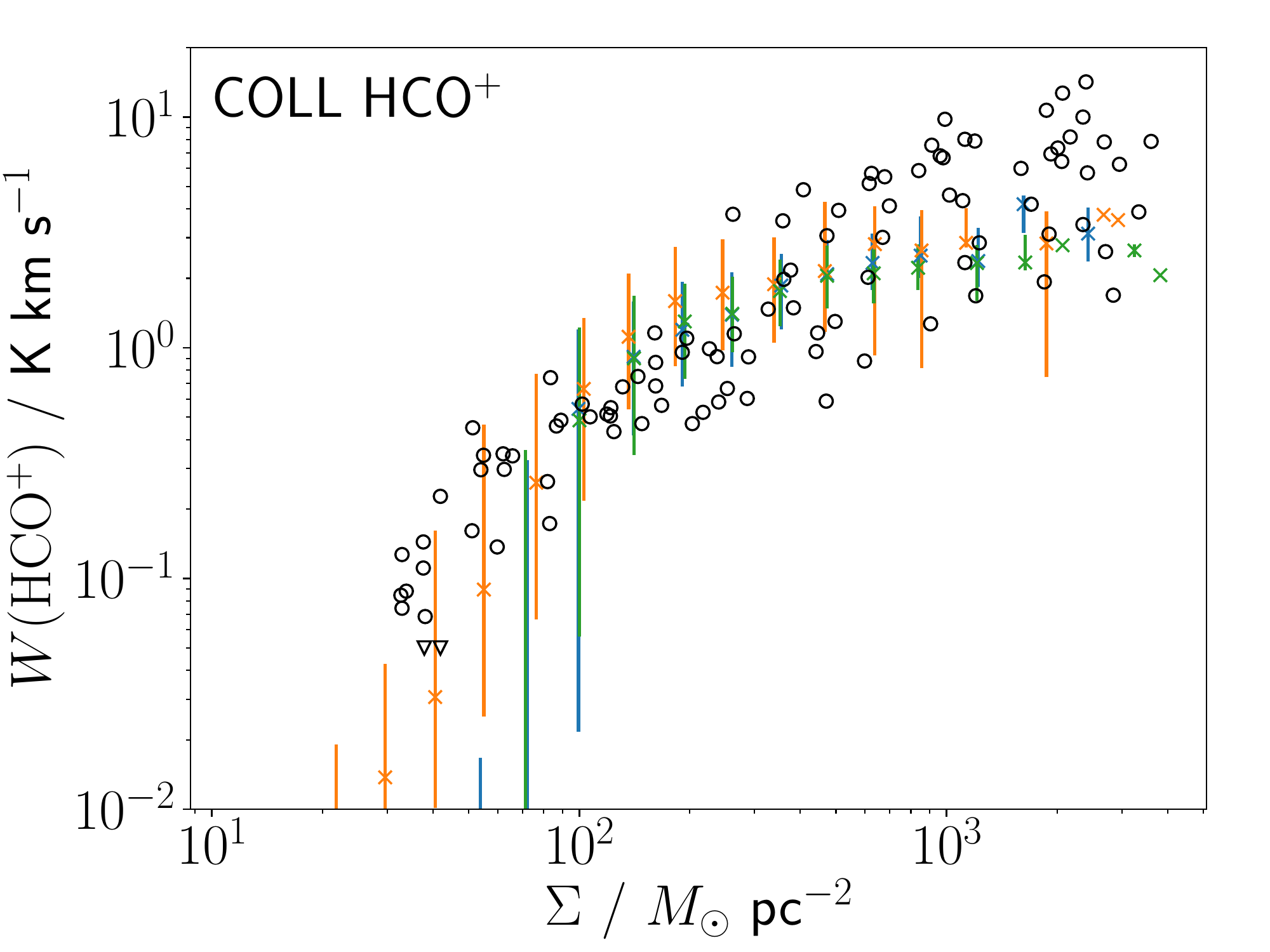}\quad
  \includegraphics[width=0.31\textwidth]{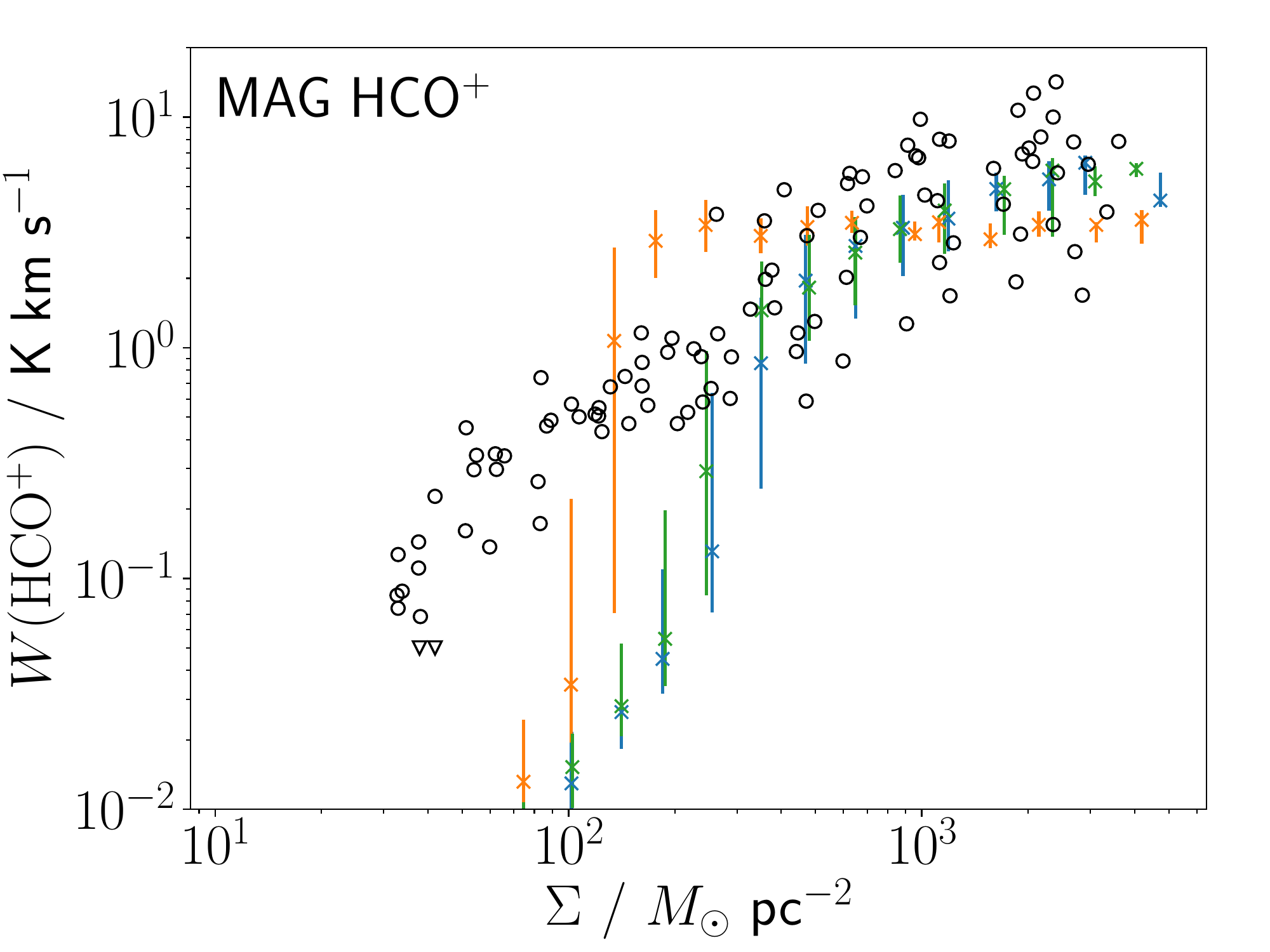}\\
  \caption{Integrated line intensity versus column density for the TURB (left), COLL (centre) and MAG (right) models, seen in the $x-y$ (orange), $x-z$ (blue) and $y-z$ (green) planes. {\it Top:} CS $J=2-1$; {\it Bottom:} HCO$^+$ $J=1-0$. Median values are shown as crosses, with the 16th and 84th percentiles as error bars. Observations from \citet{tafalla2021} are shown as circles, with those below the assumed detection limit of $0.05 \kel$ shown as triangles.}
  \label{fig:badscale}
\end{figure*}

\begin{figure}
  \centering
  \includegraphics[width=\columnwidth]{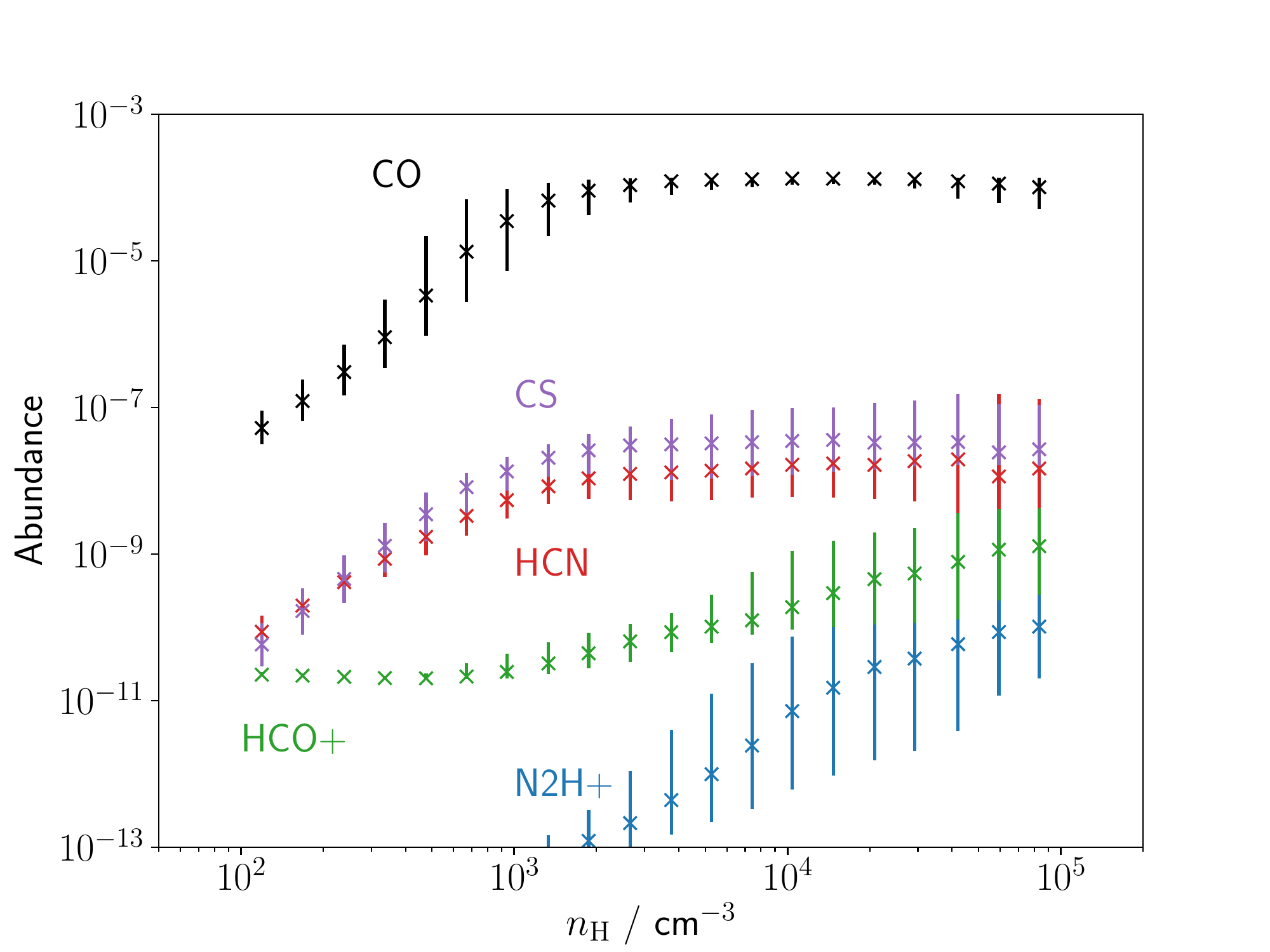}
  \caption{Molecular abundances of CO (black), HCN (red), HCO$^+$ (green), N$_2$H$^+$ (blue) and CS (purple) versus volume density for the TURB model. Median values are shown as crosses, with the 16th and 84th percentiles as error bars.}
  \label{fig:abun}
\end{figure}

We perform smoothed-particle (magneto)hydrodynamic (SPH) simulations of molecular clouds using the {\sc phantom} code \citep{price2018}. For the equation of state, we use
\begin{equation}
  T / \kel = 75 \left(\nh/10 \pcc\right)^{-1.8} + 30 \left(\nh/1000 \pcc\right)^{-0.3},
  \label{eq:eos}
\end{equation}
where $\nh$ is the number density of hydrogen nuclei in all forms, with a temperature floor of $10 \kel$ and a ceiling of $10^4 \kel$. We find that this double power-law shape provides an excellent fit to calculations of the equilibrium gas temperature between $\nh = 1$ and $1000 \pcc$ from \citet{wolfire2003}, as shown in Figure \ref{fig:eos}, and is a good approximation to the results of simulations including full non-equilibrium thermodynamics \citep[e.g.][]{smith2020}. We consider spherical clouds of mass $10^3 \msun$ and radius $3 \pc$, giving an initial mean\footnote{SPH particles are positioned randomly within the cloud volume, so the initial density is not exactly uniform.} density $\nh = 256 \pcc$, and $10^6$ SPH particles per cloud, for an SPH particle mass $m_{\rm SPH} = 10^{-3} \msun$. The clouds initially have a turbulent velocity field, with power spectrum $P(k) \propto k^{-2}$ (where $k$ is the wavenumber), a thermal mixture of compressive and solenoidal modes, and a root-mean-squared velocity of $1 \kms$, generated following \citet{lomax2015}.

We consider simulations of isolated single clouds, and of collisions between two identical clouds, displaced along the $z$ axis by $\Delta z = \pm 3 \pc$ and given an additional velocity in the $z$ direction of $v_{\rm coll} = \mp 2 \kms$. For models including magnetic fields, we use a uniform field strength of $7.6 \ug$, giving a mass-to-flux ratio of twice the critical value \citep{mouschovias1976}, and consistent with the observed column density-field strength relationship \citep{crutcher2012,liu2022,pattle2022}. We also investigate colliding models offset in the $y$ direction by $\Delta y$ to produce shearing rather than head-on collisions \citep{balfour2017}. The computational domain is a cubic box with side lengths $12 \pc$ and periodic boundary conditions, and the remainder of the volume is filled by a low-density medium with mean density $100$ times lower than that of the clouds. We run the SPH simulations until $t_{\rm end}$, when the maximum density exceeds $10^7 \pcc$. Model properties are listed in Table \ref{tab:models}.

For each of our simulations, we randomly select $10^5$ SPH particles from within the clouds, and use their density and temperature evolution as input for {\sc uclchem} \citep{holdship2017}, a time-dependent gas-grain chemical code, {using the UMIST12 reaction network \citep{mcelroy2013}}. We use elemental abundances from \citet{sembach2000}, as listed in Table \ref{tab:abun}, and assume all hydrogen is initially in the form of H$_2$. The cosmic ray ionisation rate per H$_2$ molecule is set to $1.3 \times 10^{-17} \, {\rm s^{-1}}$, and the ambient radiation field to $1.7$ Habing units \citep{habing1968}. We assume each particle is shielded by a column of gas equal to the local density multiplied by the local Jeans length,
\begin{equation}
  N_{\rm H} = \nh \frac{c_s}{\sqrt{G \rho}},
  \label{eq:shield}
\end{equation}
where $c_s$ is the sound speed, $G$ is the gravitational constant, and $\rho$ is the mass density. The visual extinction is then given by $A_{\rm V} = 6 \times 10^{-22} N_{\rm H} \, {\rm mag \, cm^2}$ \citep{bohlin1978}.

Self-shielding of H$_2$ and CO is complicated by the numerous lines involved saturating, or being shifted to different frequencies by the bulk motion of the gas. We follow \citet{glover2007} and assume the minimum possible self-shielding: the local density of the molecule multiplied by the effective particle size,
\begin{equation}
  N(i) = X(i) \, \nh \, \left(\frac{m_{\rm SPH}}{\rho}\right)^{1/3},
  \label{eq:selfshield}
\end{equation}
where $N(i)$ and $X(i)$ are the shielding column density and abundance of molecular species $i$. The suppression of the photodissociation rate due to self-shielding is then calculated following \citet{federman1979} for H$_2$ and \citet{vandishoeck1988} for CO. Our results are not sensitive to the implementation of self-shielding; if we instead use the total column density from Equation \ref{eq:shield} multiplied by the local molecular abundance, we find similar results to those obtained using Equation \ref{eq:selfshield}.

We use the three-dimensional physical and chemical structure of the clouds as input for line radiative transfer modelling with {\sc lime} \citep{brinch2010}. Each sampling point generated by {\sc lime} is assigned the properties (density, temperature, velocity, and molecular abundance) of the nearest post-processed SPH particle. We use $10^5$ sampling points, with dust properties from \citet{ossenkopf1994} and molecular data from the {\sc lamda} database \citep{schoier2005}. Our output position-position-velocity cubes have spatial resolution $0.015 \pc$ and velocity resolution $0.03 \kms$. We have made the results of the radiative transfer modelling publicly available at {http://cloudzoo.astro.cf.ac.uk/Downloads/202211/}.

As our chemical network does not distinguish between isotopes, we assume the ratios of molecular isotopologues are equal to the \citet{wilson1994} isotope values (specifically, $77$ for $^{12}$C/$^{13}$C). Under the relevant physical conditions, isotopic chemistry will alter this value by at most a factor of $2-3$ for $^{12}$CO/$^{13}$CO \citep{szucs2014}. In order to make the large number of radiative transfer calculations computationally feasible, we neglect hyperfine structure for those molecules where it is relevant, adopting {\sc lamda} datafiles using a single-level approximation. We investigate the consequences of a full treatment of hyperfine structure in Appendix \ref{sec:hfs}.

\section{Results}

Figure \ref{fig:img} shows the TURB, COLL and MAG models at $t_{\rm end}$, seen in the $x-z$ plane (i.e. with collision axis/magnetic field aligned vertically), in column density and the integrated intensity of the $^{13}$CO and HCN $J=1-0$ lines. All three models have broadly filamentary substructure, although the magnetic field has prevented fragmentation to some extent in the MAG model. This substructure is almost completely invisible in the $^{13}$CO emission, as the line becomes optically thick and saturates at relatively low densities, resulting in a much lower contrast between low- and high-density regions of the clouds. Structures therefore appear more extended than they actually are \citep{priestley2020}. HCN does not suffer from this issue, and the underlying density structures can be recovered to some extent in the TURB and MAG models. The same does not apply to the COLL simulation; the central column density peak, for example, is indiscernable in HCN emission.

We demonstrate this quantitatively in Figure \ref{fig:scale}, showing the median integrated intensities as a function of column density for the $^{13}$CO, HCN and N$_2$H$^+$ $J=1-0$ lines, compared to equivalent observational data for the Perseus molecular cloud from \citet{tafalla2021}. For all three models, the $^{13}$CO intensity rises steeply with density before plateauing beyond $\sim 100 \msun \pc^{-2}$, as the emission saturates. The model predictions are in good agreement with observation, with the exception of the MAG and COLL models seen in the $x-y$ plane (i.e. along either the collision or magnetic field axis). In these cases, there is a deficit of emission at high column density and, for the COLL model, an excess at low column density. We attribute this to the more sheet-like geometry of these two models; when seen face-on, the typical {\it volume} density for a given column density is higher than in the more extended TURB model. The line intensity is then higher at low column density due to higher collisional excitation rates, but at high column density the saturation effect becomes stronger.

The TURB model is also able to reproduce the observed HCN intensity-column density behaviour, as is the MAG model seen in the $x-z$ and $y-z$ planes. The approximately linear relationship found by \citet{tafalla2021} is recovered. Although model intensities are lower than observed, this is due to the single-level approximation for the line's hyperfine structure. When all hyperfine components are included in the radiative transfer modelling, the integrated intensity of all three components together is higher than that of the single combined line, and in good agreement with the observed values (Appendix \ref{sec:hfs}).

Regardless of viewing angle, the COLL model fails to reproduce the linear trend for HCN, predicting a plateau in intensity beginning at around $100 \msun \pc^{-2}$. A similar phenomenon happens for the MAG model in the $x-y$ plane, seen along the magnetic field axis. This occurs even with the inclusion of hyperfine structure (Appendix \ref{sec:hfs}). Unlike $^{13}$CO, HCN emission primarily originates from local overdensities which have substantial optical depths, so the line is optically thick at virtually all column densities for which there is significant emission. Figure \ref{fig:hcnratio} shows the intensity ratio of HCN to H$^{13}$CN for the TURB model,\footnote{The {\sc lamda} database does not contain collisional coefficients or energy levels for the H$^{13}$CN isotopologue, so we assume the same values as for HCN, and reduce the molecular abundance by the $^{12}$C/$^{13}$C ratio.} which is significantly lower than the abundance ratio of $77$ for column densities above $\sim 100 \msun \pc^{-2}$, signifying large departures from optically thin emission.

Increasing the integrated intensity of an optically thick line requires not just that there is more material along the line of sight, but that it has a large enough velocity dispersion for this additional material to contribute. Figure \ref{fig:sigv} shows the velocity dispersion of the gas in the $x-z$ plane for the TURB, COLL and MAG models. For the TURB and MAG models, the {\it mass-weighted} velocity dispersion increases up to a column density of $\sim 300 \msun \pc^{-2}$, beyond which it declines. However, the {\it volume-weighted} dispersion remains roughly constant. As the high-density material will become increasingly less important to the observed emission as column densities (and thus optical depths) increase, the mass-weighted dispersion becomes correspondingly less representative of the available bandwidth compared to the volume-weighted quantity (although neither is a perfect representation). The MAG model reaches lower absolute values for both quantities, leading to the lower intensity at a given column density seen in Figure \ref{fig:scale}. The COLL model has a roughly constant velocity dispersion beyond $100 \msun \pc^{-2}$, regardless of weighting, which means no additional bandwidth for emission is available, and the line intensity saturates.

The increase in velocity dispersion with column density in Figure \ref{fig:sigv} appears incompatible with the flat behaviour of line full-widths at half-maximum (FWHMs) found by \citet{tafalla2021}. However, the resulting line profiles do, in fact, reproduce this behaviour, such as the $^{13}$CO $J=1-0$ FWHMs shown in Figure \ref{fig:fwhm}, although the typical widths of $\sim 0.6 \kms$ are somewhat lower than the values seen in Perseus. The high optical depths of the lines make them unreliable tracers of the underlying gas dynamics.

The N$_2$H$^+$ $J=1-0$ line shows similar behaviour to HCN; the TURB model, and the MAG model viewed perpendicular to the field, are in reasonable agreement with the Perseus data, whereas the COLL model predicts a break in the intensity-column density relation which is not observed. However, in this case the inclusion of hyperfine structure may be sufficient to resolve the issue, although all hyperfine models in Appendix \ref{sec:hfs} appear to somewhat overpredict the N$_2$H$^+$ intensity, particularly the COLL model. Notably, all models recover the non-linear behaviour of this line, with an apparent break in the intensity scaling at around $200 \msun \pc^{-2}$. We emphasise that this was not inserted by hand; it appears to be a natural outcome of the combination of dynamics, chemistry, and radiative transfer.

The inconsistency between the predicted HCN intensity from the COLL model and the observed values appears to be quite general to simulations of colliding clouds. Figure \ref{fig:collscale} shows results from the SHEAR model, where the clouds are offset perpendicularly to the collision axis, and the COLLBPAR and COLLBPERP models, with a magnetic field parallel or perpendicular to the collision axis respectively. The SHEAR and COLLBPAR models both behave quite similarly to the COLL case (for HCN and for other lines), with a break in the intensity scaling that is not seen in the data. Only the COLLBPERP model recovers an approximately linear intensity-column density relationship, as the perpendicular magnetic field tends to resist the compression of the interacting layer and the subsequent decline in velocity dispersion. In fact, this model appears to be in better agreement with the data than the isolated MAG model in Figure \ref{fig:scale}.

The lines in Figure \ref{fig:scale} show, to some extent, agreement between the predicted and observed values. Figure \ref{fig:badscale} shows two lines where this is not the case, CS $J=2-1$ and HCO$^+$ $J=1-0$. The CS line intensity is strongly variable with viewing angle, even for the TURB model where there is no preferred direction, and while some orientations are a good match to the \citet{tafalla2021} observations, most are not. This is a radiative transfer effect; the CS column density as a function of column density is statistically identical for viewing angles where the model geometry suggests this should be the case.

Conversely, the poor agreement between predicted and observed HCO$^+$ intensities appears to be a chemical effect. Figure \ref{fig:abun} shows molecular abundances as a function of volume density for the TURB model (other models display similar qualitative behaviour). HCN, like CO, rises to a peak abundance at a density of around $10^3 \pcc$ and stays roughly constant above that point, whereas the N$_2$H$^+$ abundance is negligible at $10^3 \pcc$ and continues to rise up to $10^5 \pcc$. This causes the differences in the behaviour of the lines seen in Figure \ref{fig:scale}. In Figure \ref{fig:abun}, the HCO$^+$ abundance is more similar to N$_2$H$^+$, leading to a comparable break in the intensity scaling in Figure \ref{fig:badscale}. The observed behaviour is a linear relationship, as for HCN, suggesting that the chemistry of HCO$^+$ should have the same qualitative behaviour as this (and other) molecules.\footnote{The CS abundance behaves very similarly to that of HCN, further reinforcing that the differences in line behaviour with viewing angle are an optical depth effect.} The underlying reasons for this discrepancy are, however, beyond the scope of this paper.

\section{Discussion}

\subsection{Optically thick lines}

The near-linear relationship between intensity and column density for most lines investigated by \citet{tafalla2021} would be naturally explained by optically thin emission. However, our results suggest that lines are in fact optically thick over almost the entire column density range (Figure \ref{fig:hcnratio}), and the observed linear scaling is due to a `bandwidth' effect \citep[e.g.][]{whitworth2018} from the velocity dispersion. While it is somewhat surprising that this results in the observed near-perfect linear relationship, at least for non-colliding models, it appears to occur naturally from the combination of hydrodynamics, chemistry and radiative transfer employed. \citet{jones2021}, using a more sophisticated physical model with a simplified chemical network, also found that the HCN $J=1-0$ line is optically thick in nearly all situations where it is visible.

If lines are optically thick, it follows that there is little (if any) correlation between the column density of the molecule and the line intensity. This limits the utility of chemically post-processed simulations {\it without} radiative transfer modelling \citep[e.g.][]{gomez2022,panessa2022} for interpreting observational data. While optical depths can be estimated if a transition from a less-common isotopologue is available, this is typcally only possible at high column densities where the weaker lines are detectable. \citet{panessa2022} have no optical depth measurements for HCO$^+$, and thus no reliable column density measurements, below $\sim 10^{22} \pcs$; this is also the point (probably not coincidentally) at which their models begin disagreeing with the data. Dynamical information estimated from line profiles may also be a poor representation of the actual gas dynamics in molecular clouds (compare Figures \ref{fig:sigv} and \ref{fig:fwhm}). We suggest that more effort be invested in producing synthetic observational data from simulations, rather than attempting to extract comparable physical properties from the observations themselves.

\subsection{Cloud collisions}

Cloud collision models (with the possible exception of the COLLBPERP case; Figure \ref{fig:collscale}) appear to predict a fundamentally different relationship between line intensity and column density for several molecules, essentially because these species are concentrated in the compressed layer between clouds \citep{priestley2021} where the initial turbulent velocity dispersion has mostly dissipated. The Perseus data from \citet{tafalla2021} suggests that this cloud is much better represented as an isolated cloud, rather than one formed via a collision. Comparable observational data are available for Orion A \citep{kauffmann2017}, Orion B \citep{pety2017} and W49A \citep{barnes2020}; of these three, only Orion B has an HCN $J=1-0$ intensity scaling resembling our colliding models, which would tend to support the collision scenario proposed by \citet{enokiya2021} based on features in the CO line profiles \citep{haworth2015}.

While these `bridging features' are frequently identified in other regions \citep[e.g.][]{maity2022}, their identification is somewhat subjective, and many claimed detections bear very little resemblance to the \citet{haworth2015} diagnostic. The relatively uniform increase in HCN and other line intensities with column density in three out of four clouds suggests that these collisions may be less common than is often assumed. The differences between isolated and colliding models (Figures \ref{fig:scale} and \ref{fig:collscale}) occur for all viewing angles, unlike bridging features, which are only visible when the system is viewed along (or close to) the collision axis \citep{haworth2015,priestley2021}. This could make the line intensity scaling behaviour a useful - and independent - observational diagnostic of cloud collisions. We plan to investigate this possibility further in future work.

An important caveat here is that we have simulated collisions between {\it molecular} clouds. These are chemically quite different from collisions between lower density, initially atomic gas clouds, which, for example, do not show any CO bridging features \citep{clark2019}. Although the restriction of molecular emission to the compressed layer, and the reduction in velocity dispersion, should both still hold in the atomic case, further work is necessary to assess the impact on the line intensity scaling.

\subsection{Do simulated clouds look like real ones?}

Given the complexity of going from the gas dynamics of a molecular cloud to its line emission via its chemical evolution, there was no guarantee that the model cloud properties we selected would result in anything resembling reality. While some models, in particular the colliding ones, are in tension with the Perseus data for some lines, in general there is impressive agreement between the model predictions and the data. Perseus is only one cloud; the fact that the models reproduce the tight correlation between line intensity and column density seen in all clouds where this has been tested \citep{pety2017,kauffmann2017,barnes2020,tafalla2021} suggests that they are correctly capturing the underlying physics, particularly as it is not simply due to optically thin emission. We have not investigated the impact of varying the initial cloud parameters beyond those in Table \ref{tab:models}, but it seems highly unlikely that we have stumbled upon the `correct' values by accident. The simplified molecular cloud models in wide use in the literature appear to reproduce the observational properties of real clouds quite well.

\section{Conclusions}

We have combined MHD simulations, time-dependent chemistry, and radiative transfer models to produce synthetic molecular line observations of simulated molecular clouds. Our models reproduce the tight relationship between line intensity and column density seen in observational data. The linear scaling found for most lines by \citet{tafalla2021} in the Perseus molecular cloud is only reproduced by isolated clouds, as opposed to colliding models. All lines have high optical depths; increases in integrated intensity with column density are driven by higher velocity dispersions, not directly by the higher total column density of gas as in optically thin emission. The line emission properties of the simulated clouds are in reasonable agreement with those observed, suggesting that they are acceptable representations of reality.

%%%%%
\section*{Acknowledgements}
%%%%%
We are grateful to Mario Tafalla for a useful discussion about the observational data and its interpretation. The authors acknowledge the support of a consolidated grant (ST/K00926/1) from the UK Science and Technology Facilities Council (STFC).

%%%%%
\section*{Data Availability}
%%%%%
The data underlying this article will be shared on request. The SPH and radiative transfer data used in the analysis are publicly available at {http://cloudzoo.astro.cf.ac.uk/Downloads/202211/}.

%%%%% REFERENCES
\bibliographystyle{mnras}
\bibliography{molscale}

\appendix

\section{Hyperfine structure}
\label{sec:hfs}

\begin{figure*}
  \centering
  \includegraphics[width=0.31\textwidth]{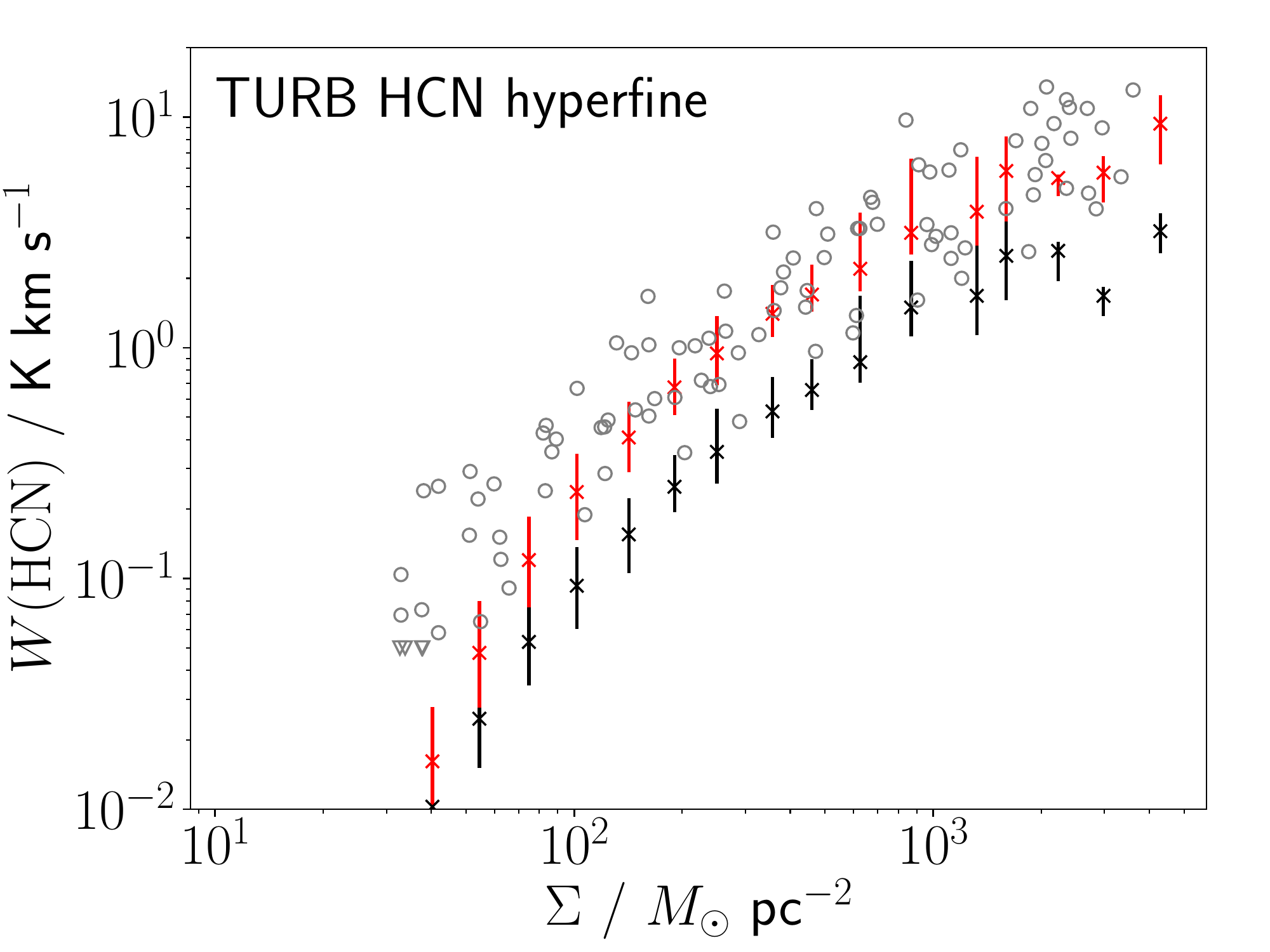}\quad
  \includegraphics[width=0.31\textwidth]{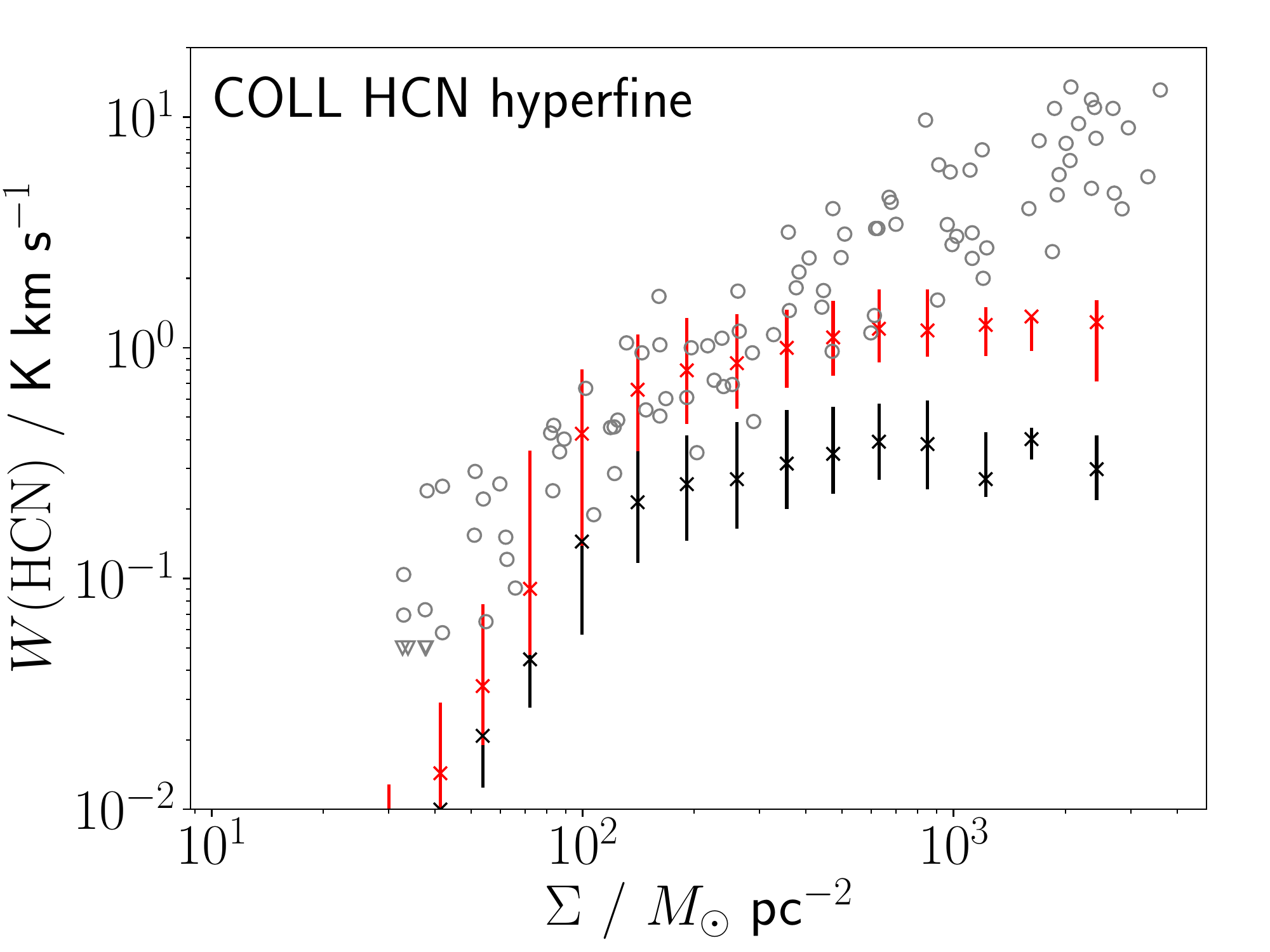}\quad
  \includegraphics[width=0.31\textwidth]{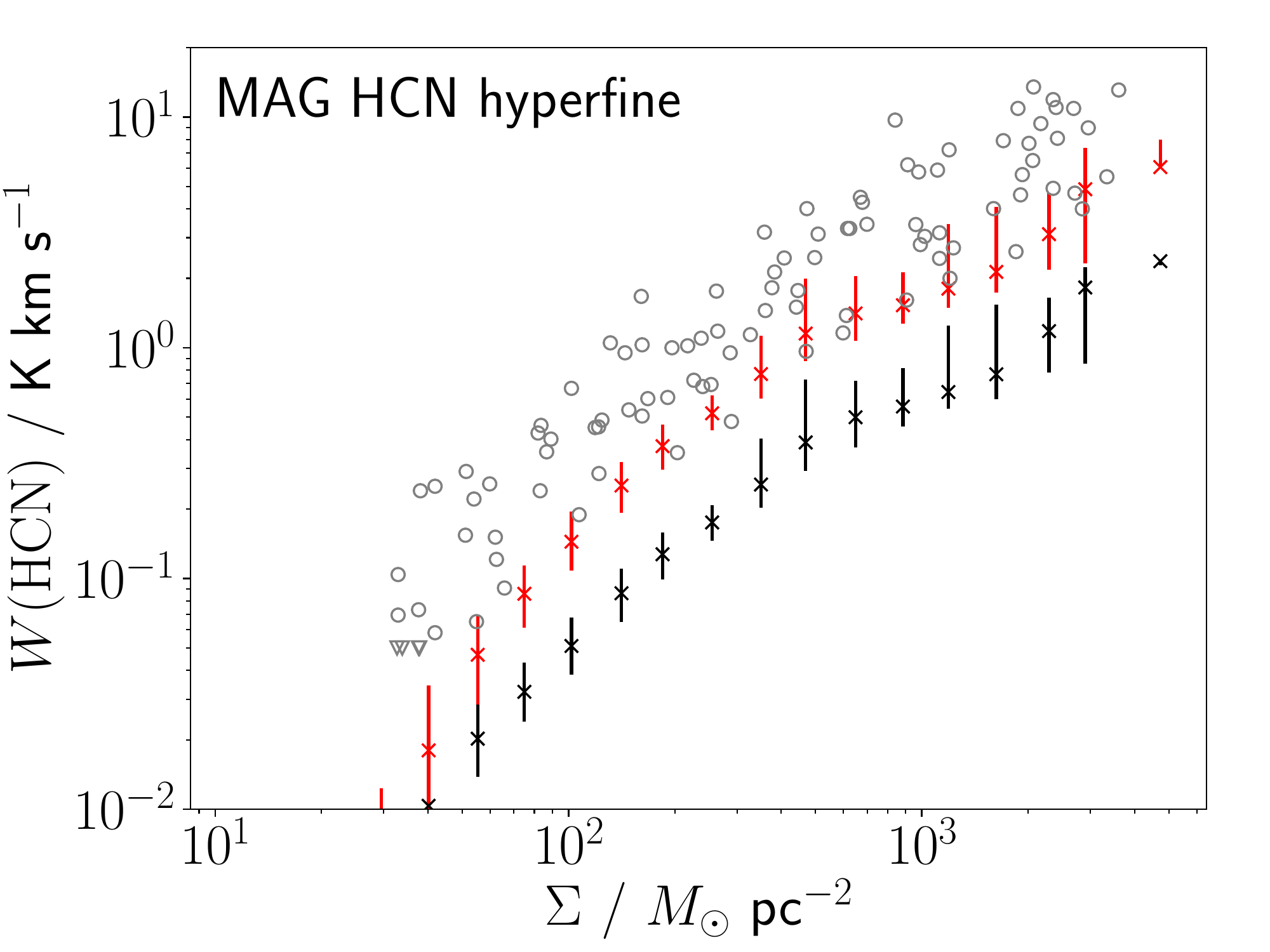}\\
  \includegraphics[width=0.31\textwidth]{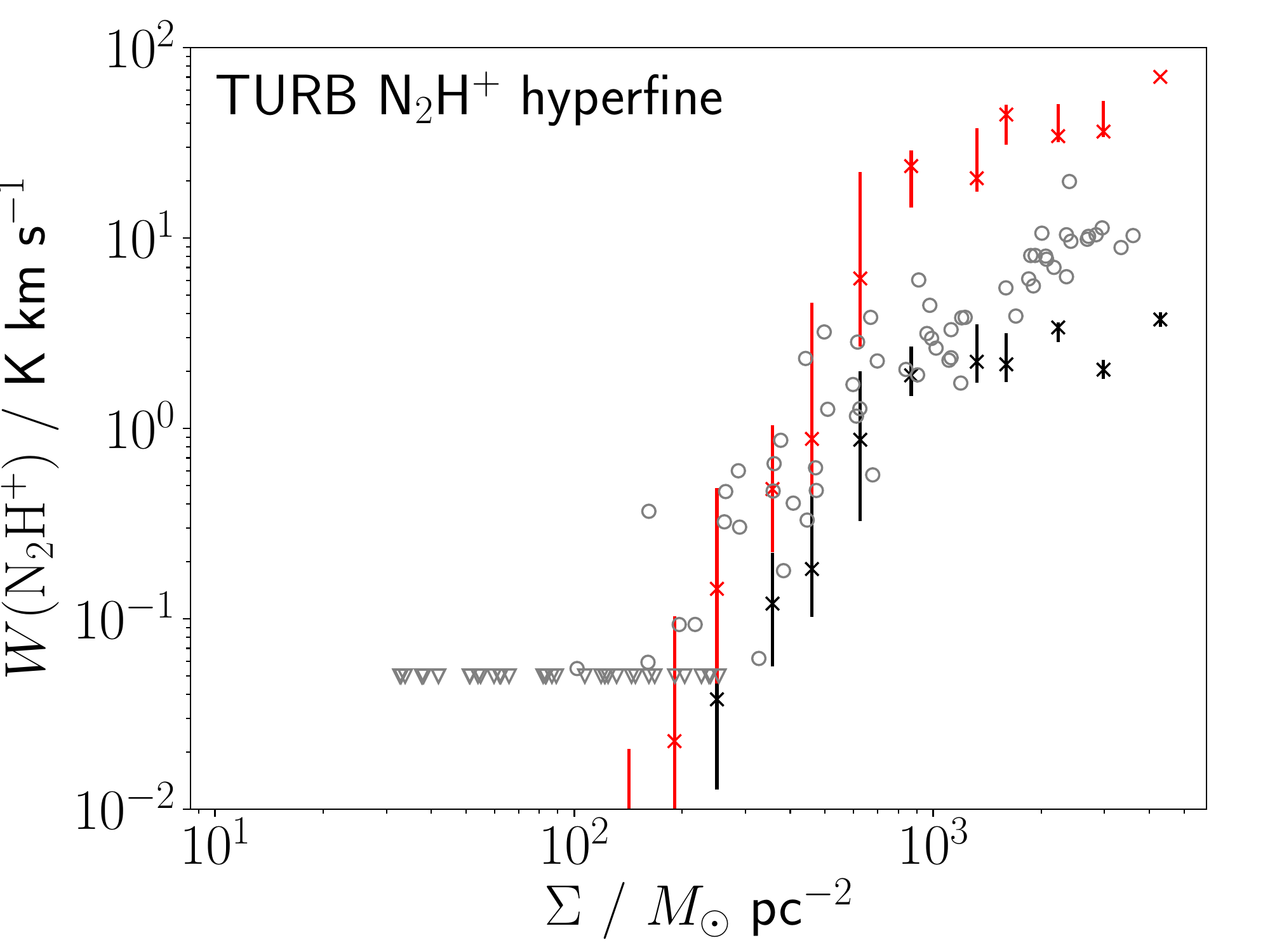}\quad
  \includegraphics[width=0.31\textwidth]{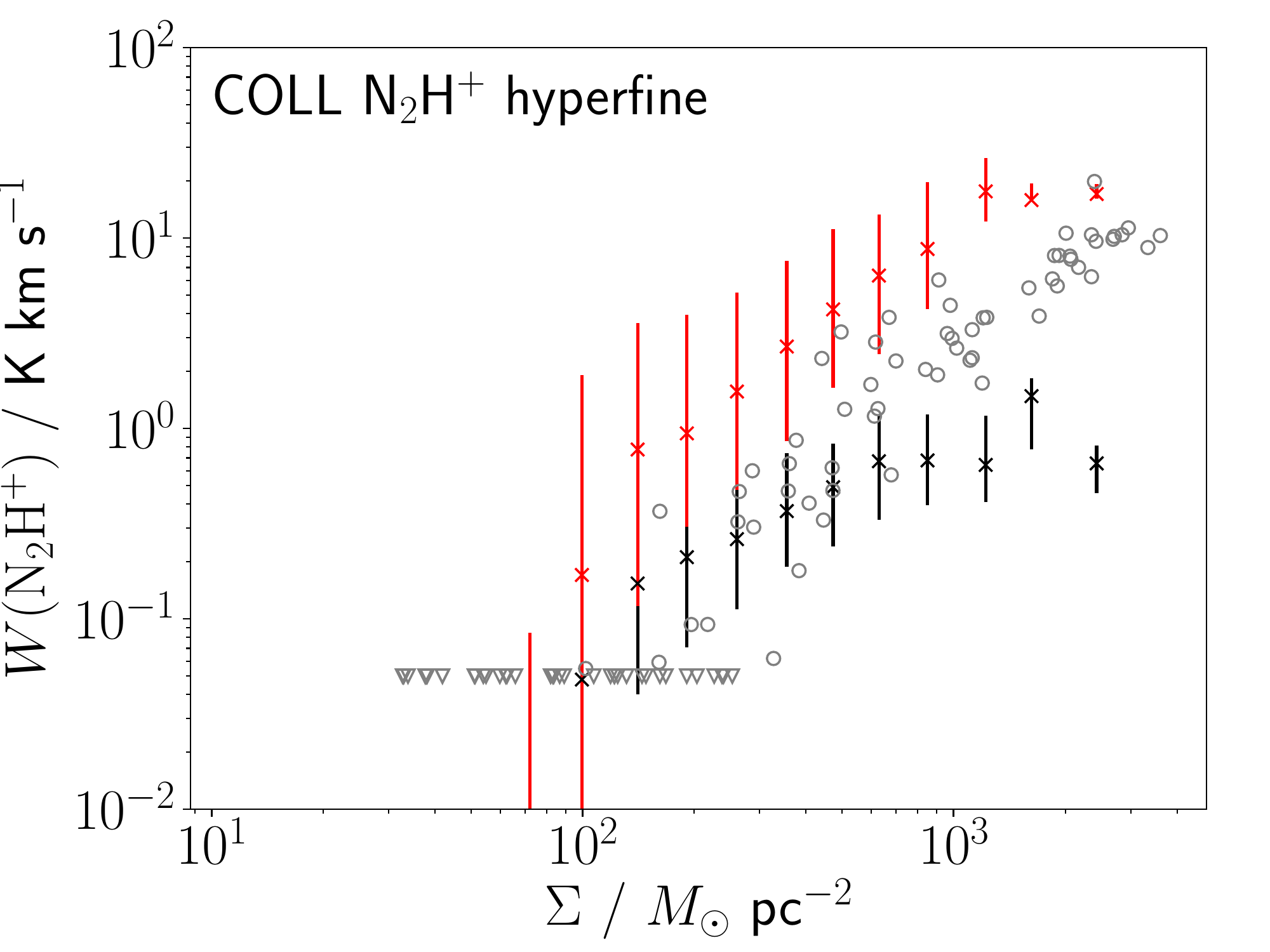}\quad
  \includegraphics[width=0.31\textwidth]{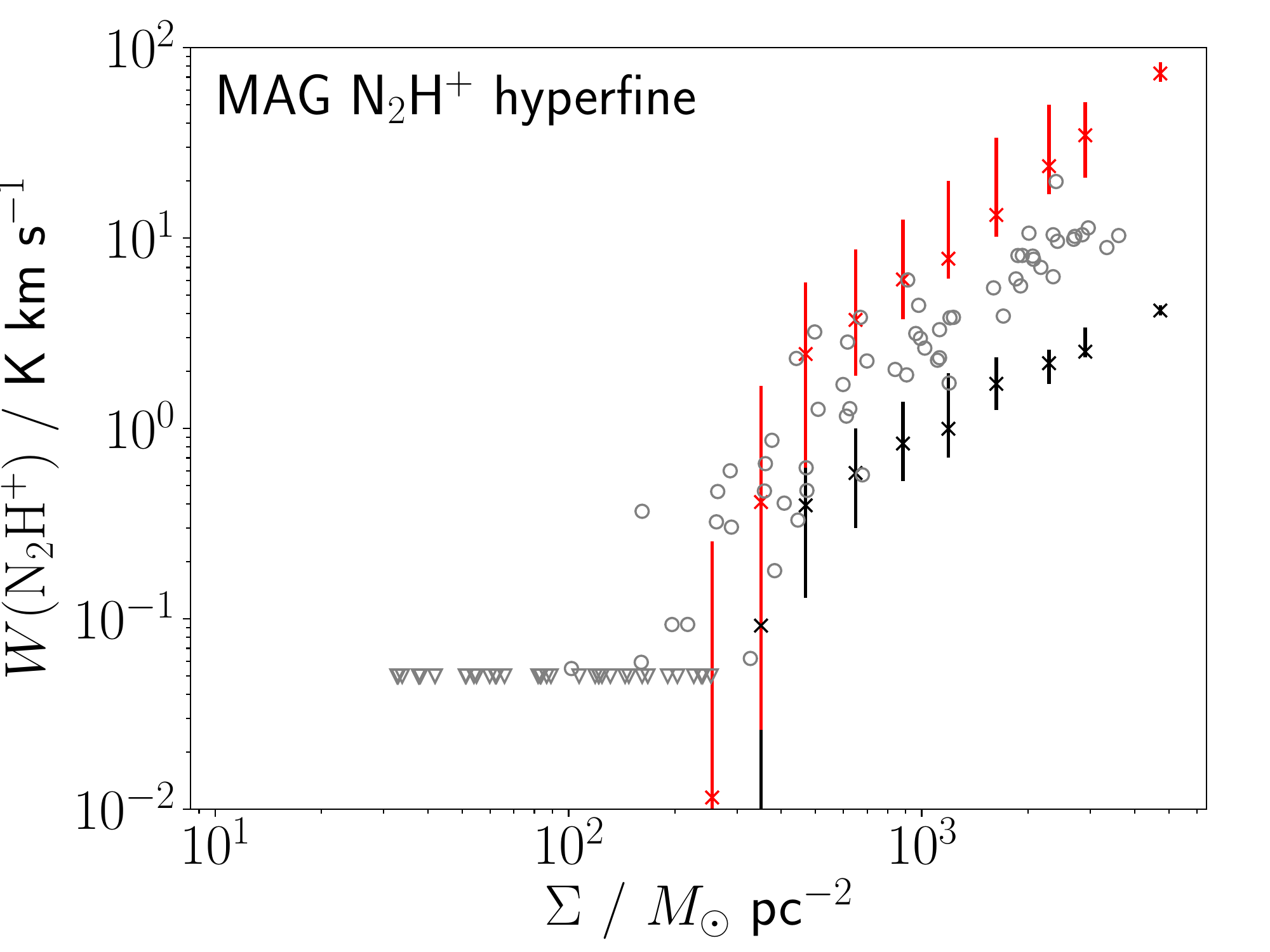}\\
  \caption{Integrated line intensity versus column density for the TURB (left), COLL (centre) and MAG (right) models seen in the $x-z$ plane, with (red) and without (black) the inclusion of hyperfine structure. {\it Top:} HCN $J=1-0$; {\it Bottom:} N$_2$H$^+$ $J=1-0$. Median values are shown as crosses, with the 16th and 84th percentiles as error bars. Observations from \citet{tafalla2021} are shown as circles, with those below the assumed detection limit of $0.05 \kel$ shown as triangles.}
  \label{fig:hfs}
\end{figure*}

Figure \ref{fig:hfs} shows the impact of hyperfine structure on the HCN and N$_2$H$^+$ $J=1-0$ line intensities for the TURB, COLL and MAG models, seen in the $x-z$ plane. For an optically thin line, the integrated intensity for the single-level approximation should be the same as the integrated intensity of all hyperfine components combined. With a significant optical depth, as in our models, the reduced opacity of the individual hyperfine components compared to the single-level approximation results in a higher line intensity at a given column density. For the HCN line, this brings the TURB and MAG model predictions into good agreement with the Perseus data, but is not sufficient to remove the plateau effect seen in the COLL model. For the N$_2$H$^+$ line, while the non-linear behaviour below $100 \msun \pc^{-2}$ is preserved, the models with hyperfine structure tend to overestimate the line emission, particularly at high column densities.

% Don't change these lines
\bsp	% typesetting comment
\label{lastpage}
\end{document}